\documentclass[12pt, fleqn, a4paper]{article} 
\usepackage{hyperref}
\hypersetup{pdfstartview = FitH,
            pdfborder    = {0 0 0.5},
            colorlinks   = false
}
 \usepackage{epsfig, subcaption}
 \usepackage{multirow}
 \usepackage{clshan-math, clshan-dm-dd, clshan-dm-ddd}
 \def \lsim {\:\raisebox{-0.7 ex}{$\stackrel{\textstyle<}{\sim}$}\:}
 \def \PeriodAa     {0      -- 365}
 \def \PeriodBa     {49.0   -- 109.0}
 \def \PeriodBb     {140.25 -- 200.25}
 \def \PeriodBc     {231.50 -- 291.50}
 \def \PeriodBd     {322.75 -- 382.75}
 \def \PeriodCa     {19.49  --  79.49}
 \def \PeriodCb     {110.74 -- 170.74}
 \def \PeriodCc     {201.99 -- 261.99}
 \def \PeriodCd     {293.24 -- 353.24}
 \def \PeriodDa     {177.66 -- 237.66}
 \def \PeriodDb     {360.16 -- 420.16}
\newcommand{\rmF}   {\rmXA{F}  {19}}
\newcommand{\rmAr}  {\rmXA{Ar} {40}}
\newcommand{\rmGe}  {\rmXA{Ge} {73}}
\newcommand{\rmXe}  {\rmXA{Xe}{129}}
\newcommand{\rmW}   {\rmXA{W} {183}}
\newcommand{\InsertSKPPlotS} [3] [8.25] {
\begin{figure} [t!]
\begin{center}
 \includegraphics [width = #1 cm] {skp-#2}
\end{center}
\caption{
 #3
}
\label{fig:#2}
\end{figure}
}
\newcommand{\InsertSKPPlotD} [4] {
\begin{figure} [t!]
\begin{center}
 \begin{subfigure} [c] {8.25 cm}
  \includegraphics [width = 8.25 cm] {skp-#1}
 \caption{}
 \end{subfigure}
 \hspace{0.1 cm}
 \begin{subfigure} [c] {8.25 cm}
  \includegraphics [width = 8.25 cm] {skp-#2}
 \caption{}
 \end{subfigure}
\end{center}
\caption{
 #4
}
\label{#3}
\end{figure}
}
\begin{document}
\thispagestyle{empty}
\begin{flushright}
 March 2021
\end{flushright}
\begin{center}
{\Large\bf
 Monte Carlo Scattering--by--Scattering Simulation of   \\ \vspace{0.2 cm}
 3-Dimensional Elastic WIMP--Nucleus Scattering Events} \\
\vspace*{0.7 cm}
 {\sc Chung-Lin Shan}                                   \\
\vspace{0.5 cm}
 {\small\it
  Preparatory Office of
  the Supporting Center for
  Taiwan Independent Researchers                        \\ \vspace{0.05 cm}
  P.O.BOX 21 National Yang Ming Chiao Tung University,
  Hsinchu City 30099, Taiwan, R.O.C.}                   \\~\\~\\
 {\it E-mail:} {\tt clshan@tir.tw}
\end{center}
\vspace{2 cm}
\begin{abstract}

 In this paper,
 as the first part of the third step of
 our study on developing data analysis procedures
 for using 3-dimensional information
 offered by directional direct Dark Matter detection experiments
 in the future,
 we present
 our double--Monte Carlo ``scattering--by--scattering'' simulation of
 the 3-dimensional elastic WIMP--nucleus scattering process,
 which can provide
 3-D velocity information
 (the magnitude,
  the direction,
  and the incoming/scattering time) of
 each incident halo WIMP
 as well as
 the recoil direction
 and the recoil energy of
 the scattered target nucleus
 in different celestial coordinate systems.
 For readers' reference,
 (animated) simulation plots
 with different WIMP masses
 and several frequently used target nuclei
 for all functionable underground laboratories
 can be found and downloaded
 on our online (interactive) demonstration webpage
 ({\tt \url{http://www.tir.tw/phys/hep/dm/amidas-2d/}}).

\end{abstract}
\clearpage
\tableofcontents
\addtocontents{toc}{}
\clearpage
\section{Introduction}

 So far
 Weakly Interacting Massive Particles (WIMPs) $\chi$
 arising in several extensions of
 the Standard Model
 of particle physics
 are still one of the most favorite
 candidates for cosmological Dark Matter (DM).
 In the last (more than)
 three decades,
 a large number of experiments has been built
 and is being planned
 to search for different WIMP candidates
 by
 direct detection of
 the scattering recoil energy
 of ambient WIMPs off target nuclei
 in low--background underground laboratory detectors
 (see Refs.~%
  \cite{SUSYDM96,
        Gaitskell04,
        Baudis12c, Baudis15,
        Drees18b,
        Schumann19,
        Baudis20}
  for reviews).

 Besides non--directional direct detection experiments
 measuring only recoil energies
 deposited in detectors,
 the ``directional'' detection of Galactic DM particles
 has been proposed more than one decade
 to be a promising experimental strategy
 for discriminating signals from backgrounds
 by using additional 3-dimensional information
 (recoil tracks and/or head--tail senses)
 of (elastic) WIMP--nucleus scattering events
 (see Refs.~%
  \cite{Ahlen09,
        Mayet11, Vahsen14, Phan15,
        Mayet16, Battat16b,
        Vahsen20,
        Vahsen21}).
 Several experimental collaborations
 investigate different detector materials and techniques
 \cite{Ikeda20, Marshall20}
 and have achieved recently great progress
 \cite{Mayet16, Battat16b,
       Vahsen20}.

\InsertSKPPlotD
 {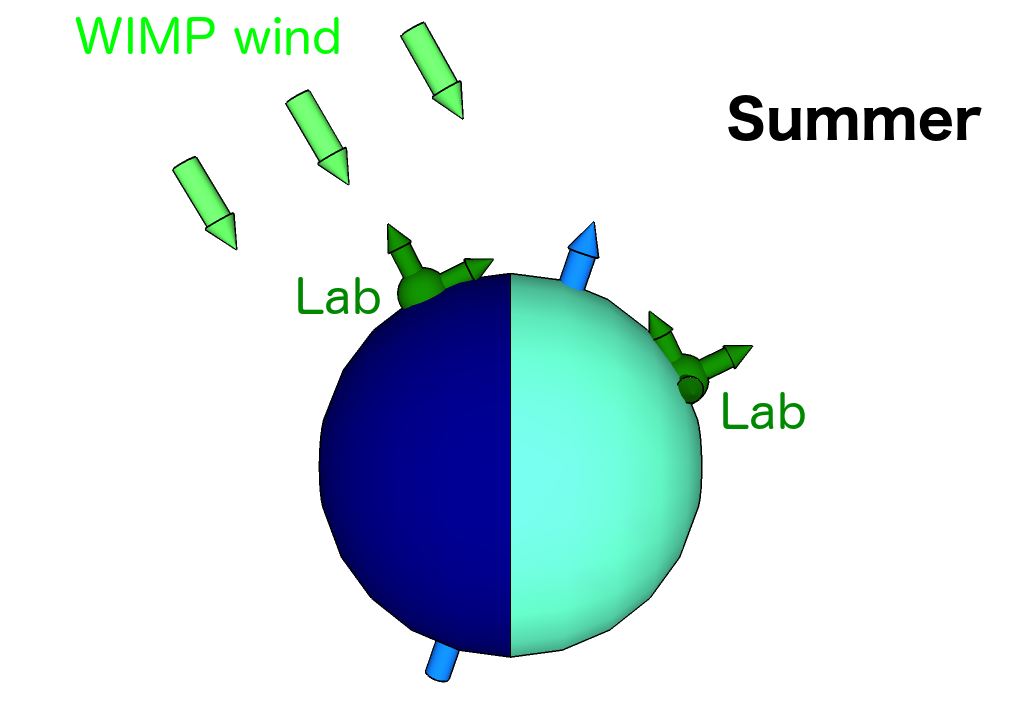}
 {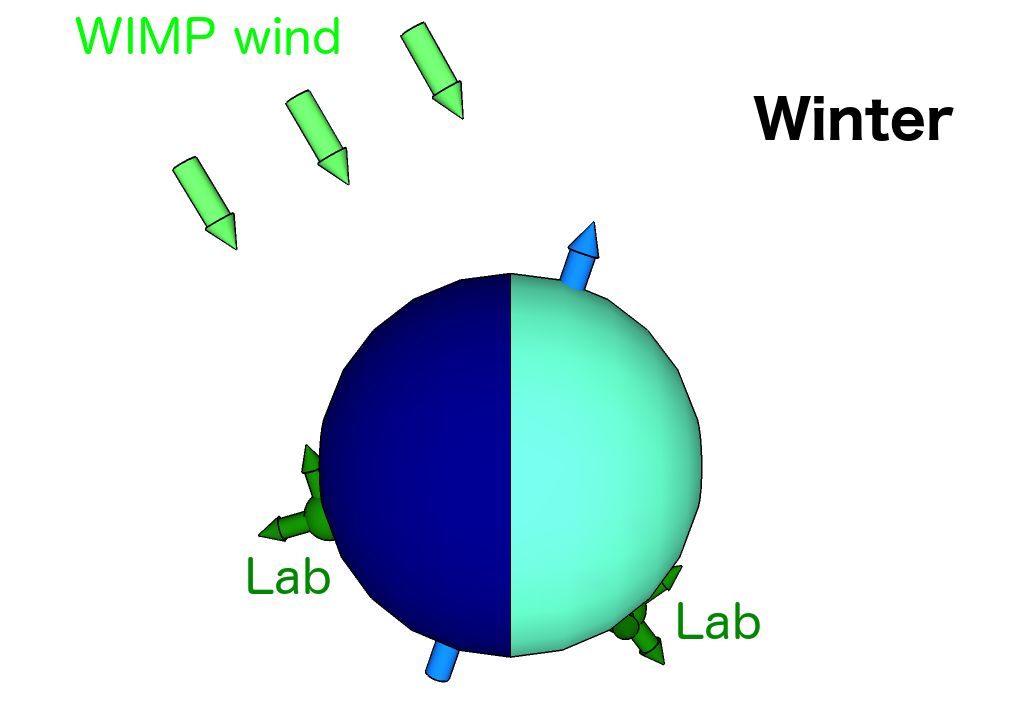}
 {fig:directional-1293-summer}
 {The basic concepts of
  directional direct Dark Matter detection:
  (a)
  the diurnal modulation of
  the (main) incident direction of halo WIMPs,
  the so--called ``directionality'' of
  the WIMP wind,
  for a laboratory located in the Northern Hemisphere
  (in Summer);
  (b)
  except of the directionality of the WIMP wind,
  the event number of WIMP signals
  observed at a laboratory located in the Southern Hemisphere
  (in Winter)
  could also have the diurnal modulation
  caused by the Earth's shielding of the WIMP flux.
  The darkened/lightened (left/right--hand) spheres
  indicate that
  the laboratory of interest is in the night/day.%
  }

 The basic concept of
 directional direct Dark Matter detection
 is based on the rotation of the Earth.
 As sketched in Figs.~\ref{fig:directional-1293-summer},
 there are two kinds of
 possible ``diurnal'' modulation of WIMP signals
 to observe:
 the diurnal modulation of
 the (main) incident direction of halo WIMPs,
 the so--called ``directionality'' of the WIMP wind,
 as well as
 that of the number (scattering rate) of WIMP events
 caused by Earth's shielding of the WIMP flux.
 Directional DM detection experiments
 aim originally hence,
 as the first step,
 to identify positive {\em modulated anisotropic} WIMP signals
 and discriminate them
 from theoretically (approximately) isotropic background events.

 As the preparation
 for our future study
 on the development of data analysis procedures
 for using and/or combining 3-D information
 offered by directional detection experiments
 to,
 e.g.,
 reconstruct the 3-dimensional WIMP velocity distribution,
 we develop step by step
 our double--Monte Carlo (MC) ``scattering--by--scattering'' simulation package
 for the {\em 3-dimensional} elastic WIMP--nucleus scattering process.
 In Ref.~\cite{DMDDD-N},
 we started with the Monte Carlo generation of
 the 3-D velocity of
 (incident) halo WIMPs
 in the Galactic coordinate system,
 including
 the magnitude,
 the direction,
 and the incoming/scattering time.
 Each generated 3-D WIMP velocity
 has then been transformed to
 the laboratory--independent
 (Ecliptic,
  Equatorial,
  and Earth)
 coordinate systems
 as well as
 to the laboratory--dependent
 (horizontal and laboratory)
 coordinate systems
 for further analyses
 \cite{DMDDD-N, DMDDD-P}.

 Now,
 we finally achieve
 the core part of our simulation package ---
 the 3-D elastic WIMP--nucleus scattering process
 and can provide
 the recoil direction and then the recoil energy of
 the WIMP--scattered target nuclei
 event by event
 in different celestial coordinate systems,
 as pseudo--data
 for future investigations on
 analysis procedures and reconstruction methods.
 In this paper,
 we focus on
 the overall simulation procedure of
 3-D elastic scattering
 by (generated and transformed) incident halo WIMPs,
 in particular,
 the validation of
 the 3-D recoil information of
 the WIMP--scattered target nuclei.
 Detailed studies on
 the angular distributions of
 the nuclear recoil direction/energy
 and
 the 3-dimensional effective velocity distribution of
 the incident WIMPs
 scattering off target nuclei
 will be presented separately
 in Refs.~\cite{DMDDD-NR} and \cite{DMDDD-fv_eff}
 respectively.

 The remainder of this paper is organized as follows.
 In Sec.~2,
 we describe
 the overall workflow of
 our double--Monte Carlo
 scattering--by--scattering simulation procedure of
 3-dimensional elastic WIMP--nucleus scattering.
 Then
 we review
 the MC generation of
 the 3-D velocity information of
 Galactic WIMPs
 as well as
 summarize
 the transformations
 between different celestial coordinate systems
 in Sec.~3.
 In Sec.~4,
 we introduce
 an incoming--WIMP coordinate system
 and describe in detail
 the validation criterion
 of our MC simulation of
 3-D elastic WIMP--nucleus scattering events.
 We summarize in Sec.~5.
 The definitions of
 all celestial coordinate systems
 used in our simulation package
 and
 the transformation matrices
 between these coordinate systems
 will be given in Appendix.

 \renewcommand{\arraystretch}{1.25}
\section{Simulation workflow}
\label{sec:workflow}
\begin{figure} [t!]
\begin{center}
 \includegraphics [width = 15.5 cm] {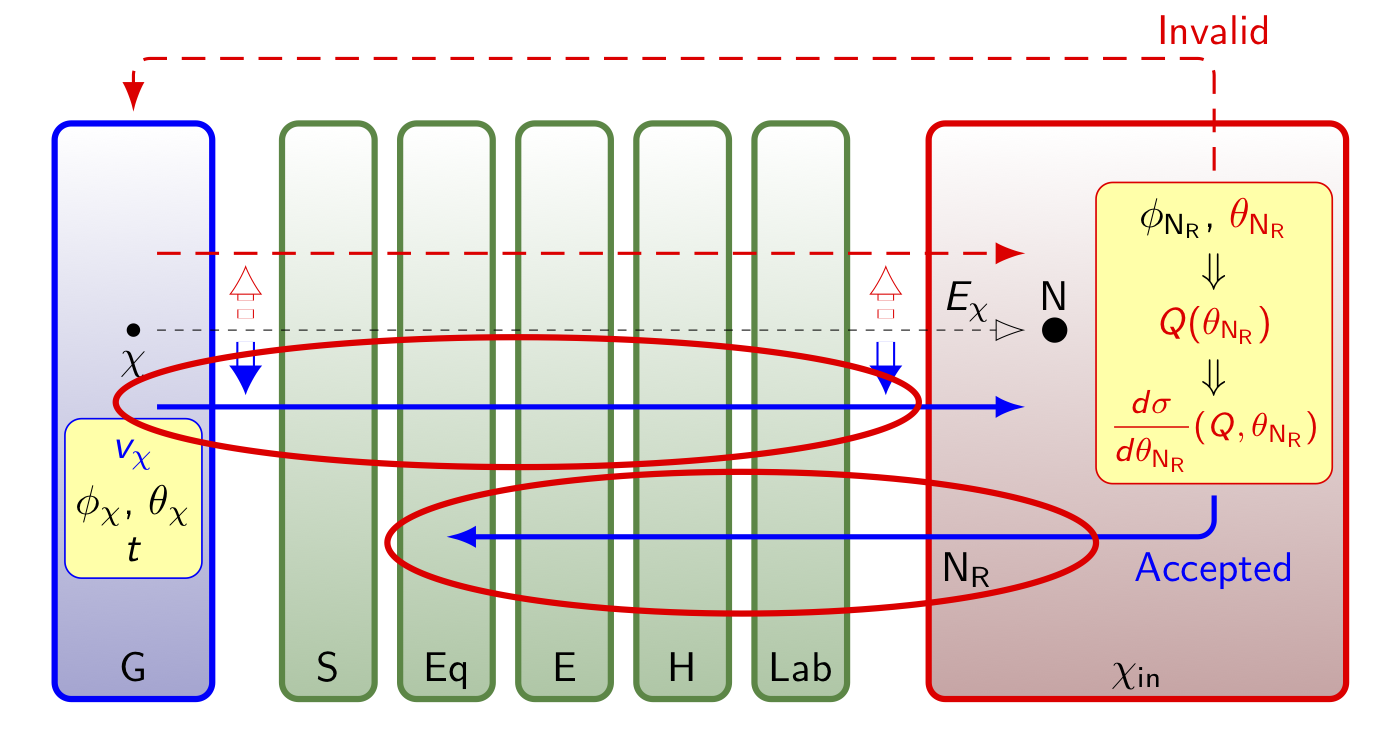}
\end{center}
\caption{
 The workflow of
 our double--Monte Carlo simulation and data analysis procedure of
 3-dimensional elastic WIMP--nucleus scattering.
 See the text for detailed descriptions.
}
\label{fig:workflow}
\end{figure}

 In this section,
 we describe
 the overall workflow of
 our double--Monte Carlo simulation and data analysis procedure of
 3-D elastic WIMP--nucleus scattering
 sketched in Fig.~\ref{fig:workflow} in detail:
\begin{enumerate}
\item
 The 3-D velocity information of incident halo WIMPs
 (the magnitude and the direction
  as well as
  the incoming/scattering time)
 is MC generated
 according to a specified model of the Dark Matter halo
 in the Galactic coordinate system
 (the blue subframe),
 which will be described
 in Sec.~\ref{sec:N-G}.
\item
 The generated 3-D WIMP velocities
 will be transformed through
 the laboratory--independent
 (Ecliptic,
  Equatorial,
  and Earth)
 coordinate systems
 as well as
 the laboratory--dependent
 (horizontal and laboratory)
 coordinate systems
 (the green subframes,
  see Sec.~\ref{sec:transformations})
 and at the end into the ``incoming--WIMP'' coordinate system
 (the red subframe),
 which definition will be given
 in Sec.~\ref{sec:XYZ_chi}.
\item
 In the incoming--WIMP coordinated system,
 the 3-D elastic WIMP--nucleus scattering process
 will also be MC simulated
 by generating
 an orientation of the scattering plane $\phiNRchi$
 and an ``equivalent'' recoil angle $\thetaNRchi$
 (defined in Sec.~\ref{sec:phi_theta_NR_chi}).
 They define the recoil direction of
 the scattered target nucleus
 and the latter,
 combined with the transformed WIMP incident velocity,
 will then be used for estimating
 the transferred recoil energy to
 the target nucleus,
 $Q(\thetaNRchi)$,
 and the differential WIMP--nucleus scattering cross section
 with respect to
 the recoil angle,
 $d\sigma / d\thetaNRchi (Q, \thetaNRchi)$,
 in our event validation criterion
 (see Sec.~\ref{sec:dsigma_dthetaNRchi}
  for details).
\item
 The orientation of
 the scattering plane $\phiNRchi$
 and the equivalent recoil angle $\thetaNRchi$
 of the {\em accepted} recoil events
 will be transformed (back)
 through all considered celestial coordinate systems
 (indicated by the lower solid blue arrow).
 All these 3-D recoil information of
 the scattered target nucleus
 accompanied with
 the corresponding recoil energy $Q$
 as well as
 the 3-D velocity of the scattering WIMP
 in different coordinate systems
 (the upper solid blue arrow)
 will be recorded
 for further analyses
 \cite{DMDDD-NR, DMDDD-fv_eff}.
\item
 For the {\em invalid} cases,
 in which
 the estimated recoil energies
 are out of the experimental measurable energy window
 or suppressed by the validation criterion,
 the generated 3-D information
 on the incident WIMP
 (the lower dashed red arrow)
 (and that on the scattered nucleus)
 will be discarded
 and the generation/validation process of
 one WIMP scattering event
 will be restarted from the Galactic coordinate system
 (the upper dashed red arrow).
\end{enumerate}
\section{MC generation and transformations of incident WIMPs}
\label{sec:}

 For the completeness and readers' reference,
 in this section,
 we review at first
 the generation of the 3-D WIMP velocity
 in the Galactic coordinate system
 and then
 summarize the transformations of
 the generated WIMP velocity
 through the laboratory--independent
 (Ecliptic,
  Equatorial,
  and Earth)
 coordinate systems
 as well as
 the laboratory--dependent
 (horizontal and laboratory)
 coordinate systems.
 While
 the definitions of
 all celestial coordinate systems
 used in our simulation package
 and
 the transformation matrices
 between these coordinate systems
 will be given in Appendix,
 discussions about our coordinate systems
 as well as
 the detailed derivations of the transformation matrices
 can be found in Ref.~\cite{DMDDD-N}.

\subsection{WIMP generation in the Galactic coordinate system}
\label{sec:N-G}

 In this subsection,
 we review briefly
 the Monte Carlo generation of
 the 3-dimensional WIMP velocity
 (the magnitude and the direction
  as well as
  the incoming/scattering time)
 in the Galactic coordinate system.

\subsubsection{Radial distribution of the 3-D WIMP velocity}
\label{sec:N_v-G}

 For generating
 the radial component (magnitude) of the 3-D WIMP velocity
 in the Galactic coordinate system,
 we consider
 the simple Maxwellian velocity distribution
 truncated at the Galactic escape velocity
 \cite{SUSYDM96}%
\footnote{
 Currently,
 as the beginning phase,
 we consider only
 the simplest model for
 (the radial and the angular components of)
 the 3-D WIMP velocity.
 In the future,
 other well--motivated halo models
 will be included.
}:
\beq
     f_{\chi, {\rm G, r}}(\vchiG)
  =  f_{1, \Gau}(\vchiG)
  =  \bbrac{  \afrac{\sqrt{\pi}}{4} \erf\afrac{\vesc}{v_0}
            - \afrac{\vesc}{2 v_0}  e^{-\vesc^2 / v_0^2}   }^{-1}
     \afrac{\vchiG^2}{v_0^3}
     e^{-\vchiG^2 / v_0^2}
\~,
\label{eqn:f1v_Gau_vesc}
\eeq
 for $v \le \vesc$,
 and $f_{\chi, {\rm G, r}}(\vchiG > \vesc) = 0$,
 where
 $v_0$
 is the Solar orbital speed around the Galactic center
 and $492~{\rm km/s} < \vesc < 587~{\rm km/s}$
 is the escape velocity from our Galaxy
 at the position of the Solar system
 \cite{RPP20AP}.

\begin{figure} [p!]
\begin{center}
 \begin{subfigure} [c] {13 cm}
  \includegraphics [width = 13 cm] {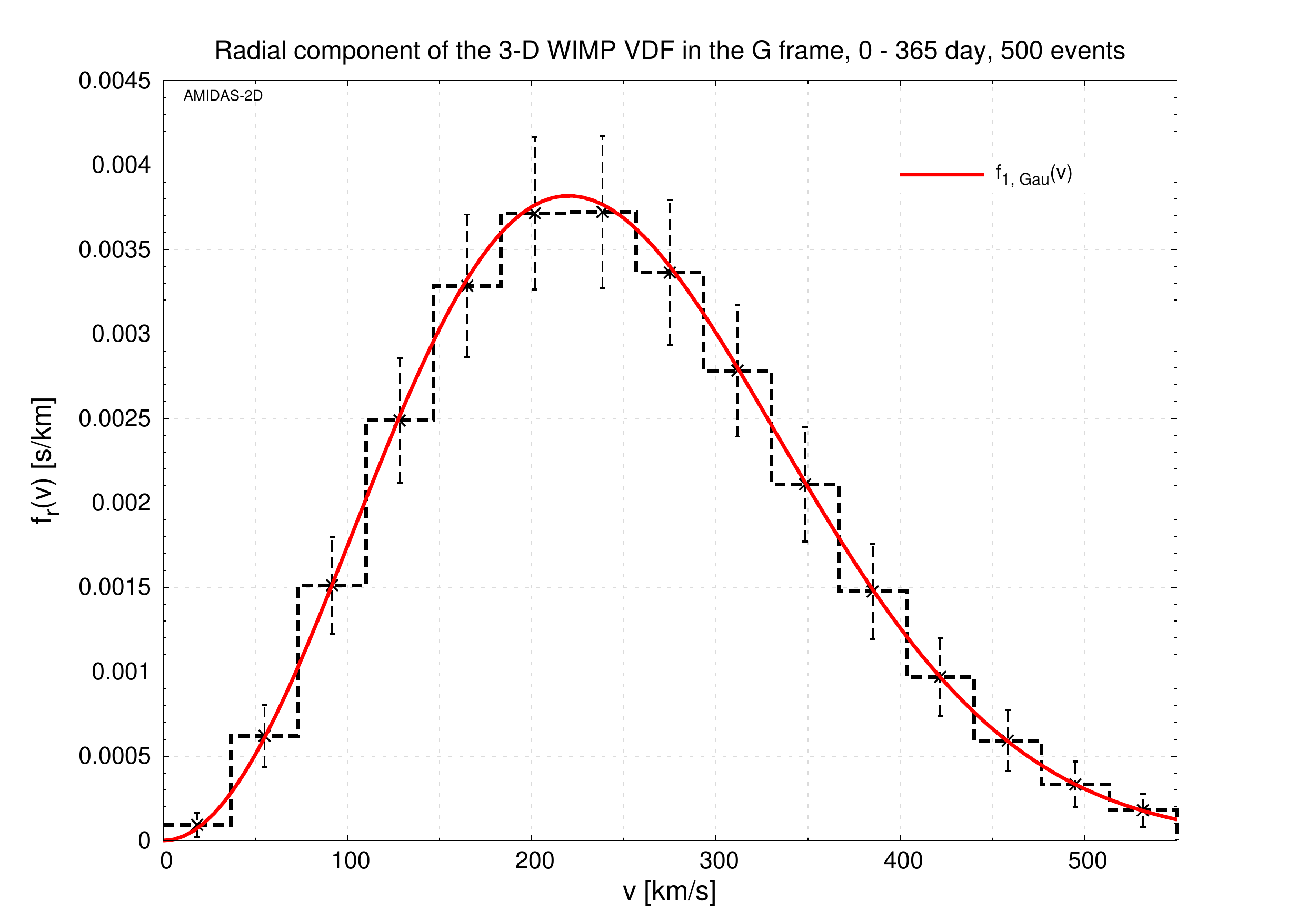}
 \caption{}
 \end{subfigure}
 \\
 \vspace{0.25 cm}
 \begin{subfigure} [c] {13 cm}
  \includegraphics [width = 13 cm] {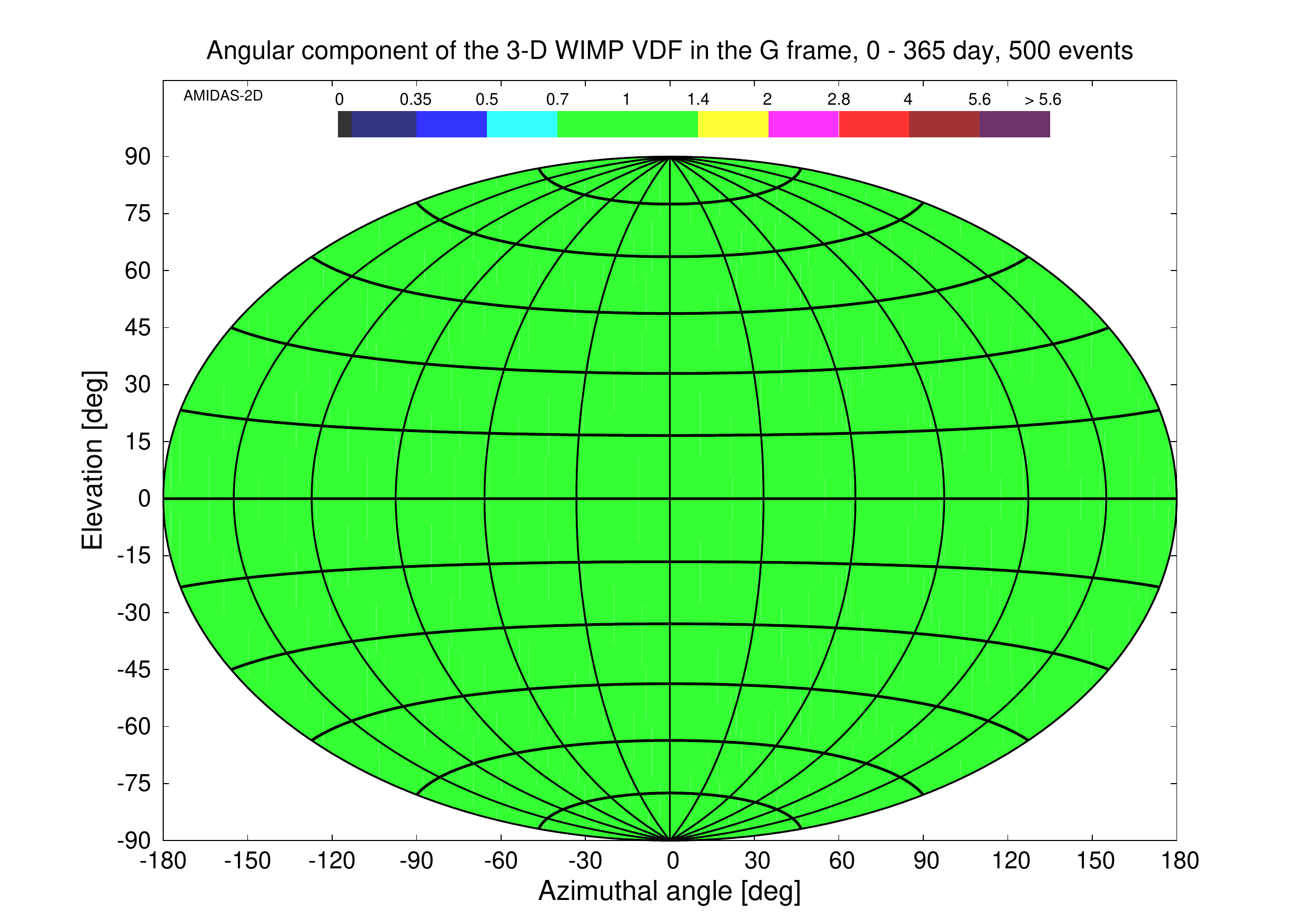}
 \caption{}
 \end{subfigure}
\end{center}
\caption{
 The radial (a) and the angular (b) components
 of the 3-D WIMP velocity
 in the Galactic coordinate system
 generated by Eq.~(\ref{eqn:f1v_Gau_vesc})
 and Eqs.~(\ref{eqn:f1v_phi_G}) and (\ref{eqn:f1v_theta_G}),
 respectively.
 The solid red curve (above)
 is the generating
 simple Maxwellian velocity distribution
 $f_{1, \Gau}(v_{\vchiG})$
 with $v_0 = 220$ km/s,
 while
 the dashed black histogram
 and the thin vertical dashed black lines
 show
 the (1$\sigma$ Poisson statistical uncertainties on the) number of
 the generated WIMP velocities.
 The horizontal color bar (below)
 indicates
 the mean value of the recorded event number
 (averaged over all simulated experiments)
 in each angular bin
 in unit of the all--sky average value
 (\mbox{500 events/144 bins $\cong$ 3.47 events/bin} here).
 See the text for further details.
}
\label{fig:N-G-0500-00000}
\end{figure}

 In Fig.~\ref{fig:N-G-0500-00000}(a),
 we show
 the radial component
 of the 3-D WIMP velocity
 in the Galactic coordinate system
 generated by Eq.~(\ref{eqn:f1v_Gau_vesc}).
 500 total events on average
 in one experiment
 (in one entire year)
 have been generated
 and binned into 15 bins.
 The solid red curve
 is the generating
 simple Maxwellian velocity distribution
 $f_{1, \Gau}(v_{\vchiG})$
 with the Solar Galactic orbital velocity $v_0 = 220$ km/s,
 while
 the dashed black histogram
 and the thin vertical dashed black lines
 show
 the (1$\sigma$ Poisson statistical uncertainties on the) number of
 the generated WIMP velocities.
 The Galactic escape velocity
 has been set as
 $\vesc = 550$ km/s.
 5,000 experiments have been simulated.

\subsubsection{Angular distribution of the 3-D WIMP velocity}
\label{sec:N_ang-G}

 Since
 the simplest model of the Galactic Dark Matter halo
 is assumed to be isothermal, spherical and isotropic,
 the angular distribution (direction) of
 the 3-D WIMP velocity
 in the Galactic coordinate system
 has been considered to be isotropic
 and thus
 the azimuthal angle $\phi$ and the elevation $\theta$
 are generated with constant probabilities:
\beq
     f_{\chi, {\rm G}, \phi}(\phichiG)
  =  1
\~,
     ~~~~ ~~~~ ~~ 
     \phichiG \in (-\pi,~\pi]
\~,
\label{eqn:f1v_phi_G}
\eeq
 and
\beq
     f_{\chi, {\rm G}, \theta}(\thetachiG)
  =  1
\~,
     ~~~~ ~~~~ ~~ 
     \thetachiG \in [-\pi / 2,~\pi / 2]
\~.
\label{eqn:f1v_theta_G}
\eeq

 In Fig.~\ref{fig:N-G-0500-00000}(b),
 we show
 the angular component
 of the 3-D WIMP velocity
 in the Galactic coordinate system
 generated by Eqs.~(\ref{eqn:f1v_phi_G}) and (\ref{eqn:f1v_theta_G}).
 500 total events on average (in one experiment
 in one entire year)
 have been binned into 12 $\times$ 12 bins
 for the azimuthal angle and the elevation,
 respectively.
 The horizontal color bar on the top of the plot
 indicates
 the mean value of the recorded event number
 (averaged over all simulated experiments)
 in each angular bin
 in unit of the all--sky average value
 (500 events/144 bins $\cong$ 3.47 events/bin here).

\subsubsection{Incoming/scattering time of 3-D WIMP--nucleus scattering events}
\label{sec:N_t-G}

 Since,
 in the Galactic point of view,
 WIMP--nucleus scattering events
 should be observed randomly and constantly,
 we consider a constant probability
 for generating the
 UTC (Coordinated Universal Time)
 incoming/scattering time of
 the recorded WIMP signals:
\beq
     f_{t}(t)
  =  1
\~,
     ~~~~ ~~~~ ~~ 
     t \in [t_{\rm start},~t_{\rm end}]
\~.
\label{eqn:f1v_t_G}
\eeq
 For example,
 for generating
 the WIMP events shown
 in Figs.~\ref{fig:N-G-0500-00000},
 the observation period has been set as
 $[t_{\rm start},~t_{\rm end}] = [0, 365~{\rm day}]$.

\subsubsection{Observation periods for annual modulations}
\label{sec:N_t-G-annual}
\InsertSKPPlotD
 {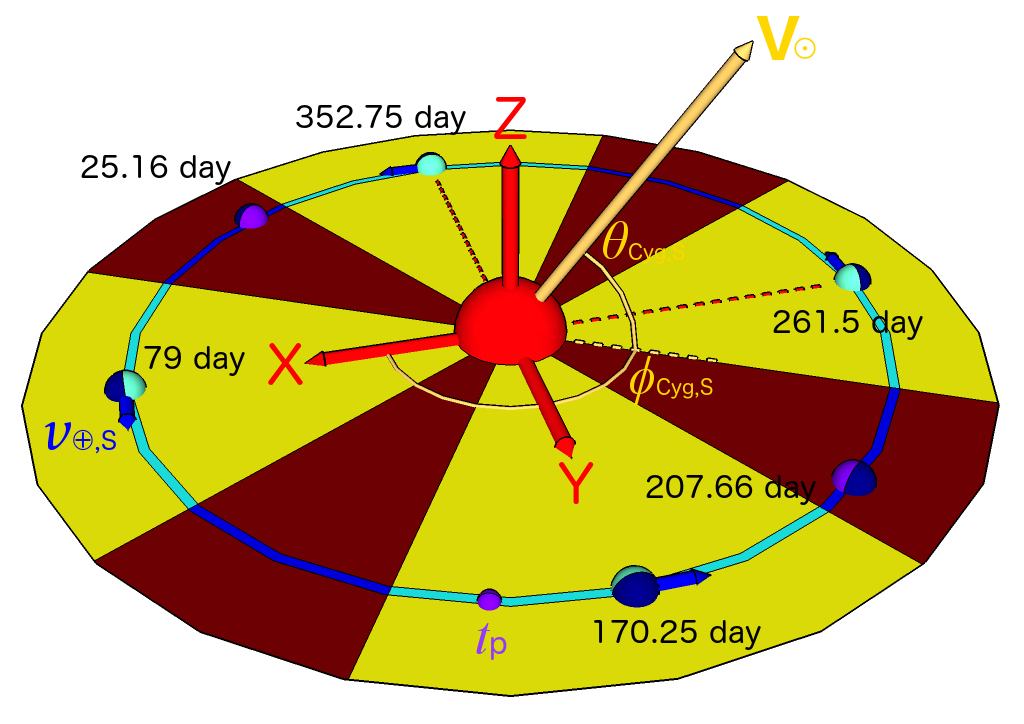}
 {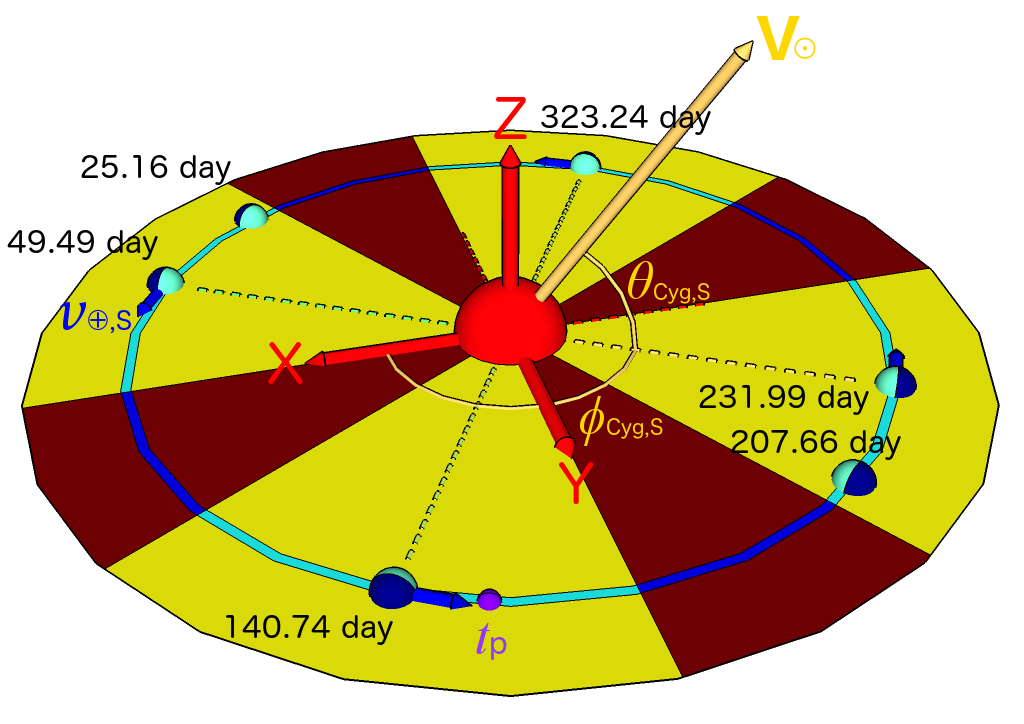}
 {fig:v_Earth_chi_S}
 {Two options for the 60-day ($\pm$ 30 days) observation periods
  (lightened areas)
  considered for demonstrating annual modulations
  (listed in Table \ref{tab:period_year}):
  (a)
  four normal seasons
  and
  (b)
  four ``advanced'' seasons.
  While
  the golden arrows indicate
  the moving direction of the Solar system
  towards the CYGNUS constellation
  with the velocity of $|\VSunG| \simeq 220$ km/s,
  the short dark--blue arrows
  (in front of the Earths)
  indicate the (average) orbital velocity of the Earth,
  $|\VEarthS| \cong 29.79$ km/s,
  on the central dates of
  the observation periods.
  Additionally,
  the Earths without the velocity arrows
  indicate the locations
  considered for demonstrating diurnal modulations
  (see also Fig.~\ref{fig:v_Earth_chi_S-20766})
  and
  the small purple points at the bottom of the sketches
  indicate the Earth's location
  around June 2nd ($t_{\rm p} = 152.5$ day)
  \cite{Freese88}.
  See Appendix \ref{appx:XYZ_G-S-Eq} and Ref.~\cite{DMDDD-N}
  for further details.%
  }
\begin{table} [t!]
\small
\begin{center}
\renewcommand{\arraystretch}{1.5}
\begin{tabular}{|| c || c | c ||}
\hline
\hline
 \makebox[5 cm][c]{Option}             &
 \makebox[5 cm][c]{Central date (day)} &
 \makebox[5 cm][c]{Period       (day)} \\
\hline
\hline
 One entire year
 & ---    & \PeriodAa \\
\hline
 \multirow{4}{*}{Four normal seasons}
 &  79.0  & \PeriodBa            \\
 & 170.25 & \PeriodBb            \\
 & 261.50 & \PeriodBc            \\
 & 352.75 & \PeriodBd\ (= 17.75) \\
\hline
 \multirow{4}{*}{Four advanced seasons}
 &  49.49 & \PeriodCa \\
 & 140.74 & \PeriodCb \\
 & 231.99 & \PeriodCc \\
 & 323.24 & \PeriodCd \\
\hline
 \multirow{2}{*}{For diurnal modulations}
 & 207.66           & \PeriodDa            \\
 & 390.16 (= 25.16) & \PeriodDb\ (= 55.16) \\
\hline
\hline
\end{tabular}
\end{center}
\caption{
 Four options
 for the observation periods
 in a 365-day year
 considered in our simulation package.
 Note that
 the last option is
 only for demonstrating diurnal modulations.
}
\label{tab:period_year}
\end{table}

 As discussed in detail
 in Ref.~\cite{DMDDD-N},
 two options
 for the observation periods
 have been considered
 for demonstrating the annual modulations of
 e.g.~%
 the angular distributions of
 the WIMP velocity (flux)
 \cite{DMDDD-N}
 and the (average) kinetic energy
 \cite{DMDDD-P}
 (sketched in Figs.~\ref{fig:v_Earth_chi_S}
  and listed in Table \ref{tab:period_year}).
 The first one is the natural choice of
 four normal seasons
 with the central dates on
 the March 21st        (79.0 day)%
\footnote{
 Note that,
 in our simulation package,
 the date of the vernal equinox
 is fixed exactly
 at the {\em end} of the May 20th (the 79th day) of
 a 365-day year
 and the few extra hours
 in an actual Solar year
 has been neglected.
},
 the June 20th         (170.25 day),
 the September 19th    (261.50 day),
 and the December 19th (352.75 day),
 respectively.

 Meanwhile,
 considering that
 the relative velocity of the Earth
 to the Galactic Dark Matter halo
 should be the maximum (minimum),
 when its orbital velocity is (anti--)parallel to
 the projection of the direction of the Solar movement
 on the Ecliptic plane
 around the 21st of May (140.74 day)
 (the 20th of November, 323.24 day),
 the second option is
 four ``advanced'' seasons
 ($\sim$ 30 days earlier)
 with the central dates on
 the February 19th     (49.49 day),
 the May 21st          (\mbox{140.74 day}),
 the August 20th       (231.99 day),
 and the November 20th (323.24 day),
 respectively.
 For each season of these two options,
 we considered
 a 60-day ($\pm$ 30 days) observation period
 and
 each pair of the corresponding season
 has thus an overlap of around 30 days.

\subsubsection{Daily shifts for diurnal modulations}
\label{sec:N_t-G-diurnal}
\begin{table} [b!]
\small
\begin{center}
\renewcommand{\arraystretch}{1.5}
\begin{tabular}{|| c || c | c ||}
\hline
\hline
 \makebox[5 cm][c]{Option}              &
 \makebox[5 cm][c]{Central time (hour)} &
 \makebox[5 cm][c]{Interval     (hour)} \\
\hline
\hline
 One entire day
 & --- &  0 -- 24 \\
\hline
 \multirow{4}{*}{Four daily shifts}
 &  0  &  0 --  2, 22 -- 24 \\
 &  6  &  4 --  8 \\
 & 12  & 10 -- 14 \\
 & 18  & 16 -- 20 \\
\hline
\hline
\end{tabular}
\end{center}
\caption{
 Two options
 for the observation intervals
 in a 24-hour day
 considered in our simulation package.
}
\label{tab:interval_day}
\end{table}

 For demonstrating
 (the originally proposed) diurnal modulations,
 we considered four observation intervals of
 4 hours ($\pm 2$ hours)
 at the central ({\em local}, not the UTC) times of
  0,
  6,
 12, and
 18 o'clock,
 respectively
 (listed in Table \ref{tab:interval_day}),
 in the 60-day periods
 centered on
 the January 26th  (25.16 = 390.16 day)
 and the July 27th (207.66 day),
 respectively
 (sketched in Figs.~\ref{fig:v_Earth_chi_S}
  and \ref{fig:v_Earth_chi_S-20766},
  see Ref.~\cite{DMDDD-N}
  for details).

\InsertSKPPlotS
 {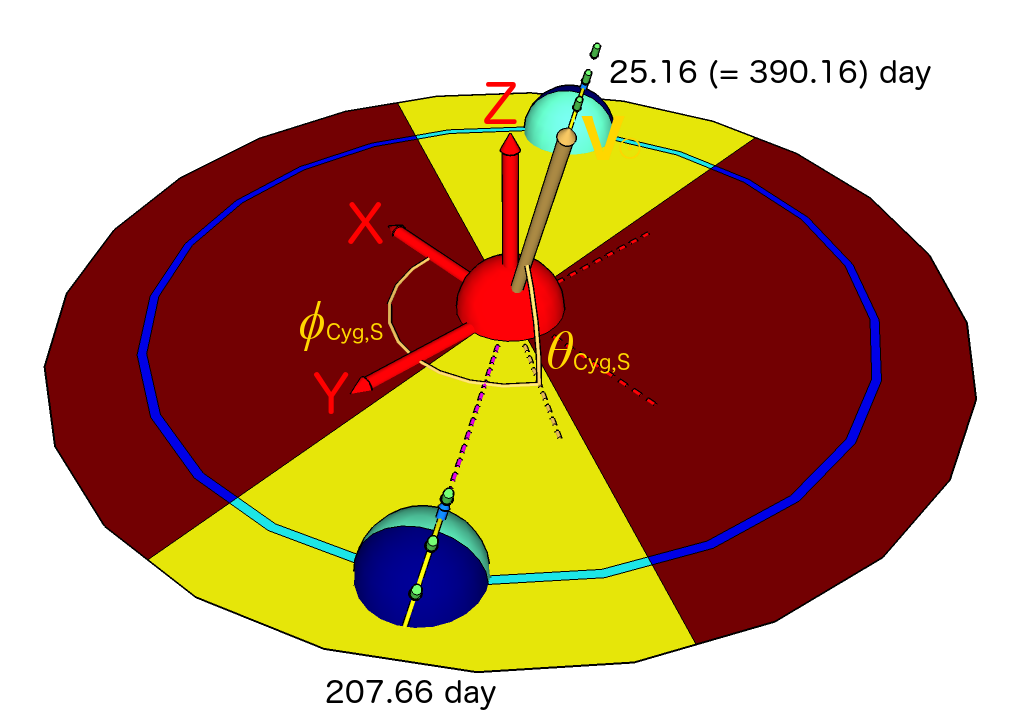}
 {Two options for the 60-day ($\pm$ 30 days) observation periods
  (lightened areas)
  considered for demonstrating diurnal modulations
  (listed in Table \ref{tab:period_year}),
  on which
  the (light--green) theoretical main direction of
  incident halo WIMPs
  points straightly to the (yellow) Prime Meridian
  in the night
  or the day.
  See Appendix \ref{appx:XYZ_G-S-Eq} and Ref.~\cite{DMDDD-N}
  for further details.%
  }
\subsection{Transformations of the 3-D velocity}
\label{sec:transformations}

 In this subsection,
 we summarize the transformations of
 the generated 3-D WIMP velocity
 as well as
 the recoil direction of
 the scattered target nucleus
 between different celestial coordinate systems.
 The analytic and/or numerical forms of
 the needed transformation matrices
 will be summarized in Appendix.

\subsubsection{Between the Galactic and the Ecliptic coordinate systems}
\label{sec:tr-G-S}

 In our simulation package,
 the 3-D WIMP velocity
 in the Ecliptic coordinate system
 ${\bf v}_{\chi, {\rm S}}$
 transformed from the generated 3-D velocity
 in the Galactic coordinate system
\(
     {\bf v}_{\chi, {\rm G}}(v_{\rm \chi, G, x}, v_{\rm \chi, G, y}, v_{\rm \chi, G, z})
  =  {\bf v}_{\chi, {\rm G}}(\vchiG, \phichiG, \thetachiG)
\)
 can be given by
\cheqna
\beq
     {\bf v}_{\chi, {\rm S}}
  =  \MaGS ({\bf v}_{\chi, {\rm G}} - \VSunG)
\~,
\label{eqn:VchiG->VchiS}
\eeq
 with the transformation matrix $\MaGS$
 given in Eq.~(\ref{eqn:Ma_G_S})
 and
\cheqn
\beq
         \VSunG
 \simeq  \left[\begin{array}{c}
                  33.58~{\rm km/s} \\
                 217.41~{\rm km/s} \\
                   2.32~{\rm km/s} \\
               \end{array}\right]_{\rm G}
\label{eqn:V_Sun_G}
\eeq
 is the moving velocity of the Solar system
 (towards the CYGNUS constellation)%
\footnote{
 Note that,
 in our simulation package,
 the Solar moving velocity $\VSunG$
 is constant
 and
 the Ecliptic coordinate system
 only moves approximately linearly
 with $\vSunG = |\VSunG| \simeq 220$ km/s;
 its tiny Galactic orbital rotation
 is considered to be imperceptible.
}
 in the Galactic coordinate system
 \cite{DMDDD-N}.
 Conversely,
 for the recoil direction of
 the WIMP--scattered target nucleus,
 we can use
\cheqnNx{-2}{b}
\beq
     {\bf v}_{\rm N_R, G}
  =  \MaSG {\bf v}_{\rm N_R, S} + \VSunG
\~,
\label{eqn:VNRS->VNRG}
\eeq
\cheqnN{1}
 where the transformation matrix $\MaSG$
 is given in Eq.~(\ref{eqn:Ma_S_G}).

\subsubsection{Between the Ecliptic and the Equatorial coordinate systems}
\label{sec:tr-S-Eq}

 Similar to Eqs.~(\ref{eqn:VchiG->VchiS})
 and (\ref{eqn:VNRS->VNRG}),
 the 3-D WIMP velocity
 in the Equatorial coordinate system
 ${\bf v}_{\chi, {\rm Eq}}$
 transformed from the transformed 3-D velocity
 in the Ecliptic coordinate system
 ${\bf v}_{\chi, {\rm S}}$
 can be given by
\cheqna
\beq
     {\bf v}_{\chi, {\rm Eq}}
  =  \MaSEq \bBig{{\bf v}_{\chi, {\rm S}} - \VEarthS(t)}
\~,
\label{eqn:VchiS->VchiEq}
\eeq
 where $t$ is the UTC incoming/scattering time of
 the WIMP event,
 the transformation matrix $\MaSEq$
 is given in Eq.~(\ref{eqn:Ma_S_Eq})
 and
\cheqn
\beq
     \VEarthS(t)
  =  \vEarthS
     \left[\renewcommand{\arraystretch}{1.5}
           \begin{array}{c}
            -\sin(\psiyear(t)) \\
             \cos(\psiyear(t)) \\
              0                     \\
           \end{array}\right]_{\rm S}
\~,
\label{eqn:V_Earth_S}
\eeq
 is the time--dependent
 Earth's orbital velocity around the Sun
 in the Ecliptic coordinate system.
 Here
 the Earth's orbital speed
 can be estimated as
 \cite{RPP20AP}%
\footnote{
 Note that,
 in our simulation package,
 the Earth's orbit around the Sun
 has been assumed to be perfectly circular
 on the Ecliptic plane
 and the orbital speed is thus a constant.
}
\beq
         \vEarthS
  \cong  29.79~{\rm km/s}
\~,
\label{eqn:v_Earth_S}
\eeq
 and
 the angle
 swept by the connection
 between the Solar and the Earth's centers
 from the day of the vernal equinox
 (the 79th day)
 can be expresses by
\beq
         \psiyear(t)
 \equiv  \frac{2 \pi}{365} \bBig{(t - \tPM) - 79.0}
\~,
\label{eqn:psi_yr}
\eeq
 where
 $\tPM$ indicates
 the fractional part of
 the UTC incoming/scattering time $t$
 in unit of day.
 Conversely,
 for the recoil direction of
 the WIMP--scattered target nucleus,
 we have
\cheqnNx{-4}{b}
\beq
     {\bf v}_{\rm N_R, S}
  =  \MaEqS {\bf v}_{\rm N_R, Eq} + \VEarthS(t)
\~,
\label{eqn:VNREq->VNRS}
\eeq
\cheqnN{3}
 with the transformation matrix $\MaEqS$
 given in Eq.~(\ref{eqn:Ma_Eq_S}).

\subsubsection{Between the Equatorial and the Earth coordinate systems}
\label{sec:tr-Eq-E}

 In our simulation package,
 the transformations of
 the 3-D (WIMP) velocity
 at the incoming/scattering time $t$
 between the Equatorial
 and the Earth coordinate systems
 are pure rotations,
 which can be given by
\cheqna
\beq
     {\bf v}_{\chi, {\rm E}}
  =  \MaEqE(t) {\bf v}_{\chi, {\rm Eq}}
\~,
\label{eqn:VchiEq->VchiE}
\eeq
 and,
 conversely,
 one has
\cheqnb
\beq
     {\bf v}_{\rm N_R, Eq}
  =  \MaEEq(t) {\bf v}_{\rm N_R, E}
\~,
\label{eqn:VNRE->VNREq}
\eeq
\cheqn
 with the time--dependent transformation matrices
 $\MaEqE(t)$ and $\MaEEq(t)$
 given in Eqs.~(\ref{eqn:Ma_Eq_E}) and (\ref{eqn:Ma_E_Eq}),
 respectively.

\subsubsection{Between the Earth and the horizontal coordinate systems}
\label{sec:tr-E-H}

 By definition,
 the transformations of
 the 3-D (WIMP) velocity
 (at the UTC incoming/scattering time $t$)
 between the Earth
 and the horizontal coordinate systems
 are pure {\em time--independent} rotations,
 which can be given by
\cheqna
\beq
     {\bf v}_{\chi, {\rm H}}
  =  \MaEH(\phiLab, \thetaLab) {\bf v}_{\chi, {\rm E}}
\~,
\label{eqn:VchiE->VchiH}
\eeq
 and,
 conversely,
\cheqnb
\beq
     {\bf v}_{\rm N_R, E}
  =  \MaHE(\phiLab, \thetaLab) {\bf v}_{\rm N_R, H}
\~,
\label{eqn:VNRH->VNRE}
\eeq
\cheqn
 where the transformation matrices
 $\MaEH(\phiLab, \thetaLab)$ and $\MaHE(\phiLab, \thetaLab)$
 depending {\em only} on the longitude and the latitude of
 the location of the considered laboratory
 $(\phiLab, \thetaLab)$
 are given in Eqs.~(\ref{eqn:Ma_E_H}) and (\ref{eqn:Ma_H_E}).

\subsubsection{Between the horizontal and the laboratory coordinate systems}
\label{sec:tr-H-Lab}

 Similar to Eqs.~(\ref{eqn:VchiE->VchiH})
 and (\ref{eqn:VNRH->VNRE}),
 the transformations (pure rotations) of
 the 3-D (WIMP) velocity
 at the incoming/scattering time $t$
 between the horizontal
 and the laboratory coordinate systems
 can be given by
\cheqna
\beq
     {\bf v}_{\chi, {\rm Lab}}
  =  \MaHLab(t, \phiLab, \thetaLab) {\bf v}_{\chi, {\rm H}}
\~,
\label{eqn:VchiH->VchiLab}
\eeq
 and,
 conversely,
\cheqnb
\beq
     {\bf v}_{\rm N_R, H}
  =  \MaLabH(t, \phiLab, \thetaLab) {\bf v}_{\rm N_R, Lab}
\~,
\label{eqn:VNRLab->VNRH}
\eeq
\cheqn
 where the transformation matrices
 $\MaHLab(t, \phiLab, \thetaLab)$ and $\MaLabH(t, \phiLab, \thetaLab)$
 depending not only on the longitude and the latitude of
 the laboratory location
 $(\phiLab, \thetaLab)$
 but also on the incoming/scattering time $t$ ($\tPM$)
 are given in Eqs.~(\ref{eqn:Ma_H_Lab}) and (\ref{eqn:Ma_Lab_H}).

\subsubsection{Between the laboratory and the incoming--WIMP coordinate systems}
\label{sec:tr-Lab-chi}

 Finally,
 for the transformation (pure rotation) of
 the recoil direction of
 the WIMP--scattered target nucleus
 generated in the incoming--WIMP coordinate system
 to the laboratory coordinate system,
 one has
\beq
     {\bf v}_{\rm N_R, Lab}
  =  \MachiLab(\phichiLab, \thetachiLab) {\bf v}_{\rm N_R, \chi_{in}}
\~,
\label{eqn:VNRchi->VNRLab}
\eeq
 where the transformation matrix
 $\MachiLab(\phichiLab, \thetachiLab)$
 depending on the azimuthal angle $\phichiLab$
 and the elevation $\thetachiLab$ of
 the incident direction of the scattering WIMP
 measured in the laboratory coordinate system
 is given in Eq.~(\ref{eqn:Ma_chi_Lab}).

\section{MC generation of 3-D elastic WIMP--nucleus scattering events}
\label{sec:3D-WIMP-N}

 As described in Sec.~\ref{sec:workflow},
 each generated 3-D WIMP velocity
 will be transformed through
 different celestial coordinate systems
 and at the end into
 the ``incoming--WIMP'' ($\chiin$) coordinate system.
 In this section,
 we describe then the core part of
 our simulation procedure:
 the generation of
 3-D elastic WIMP--nucleus scattering events
 in the incoming--WIMP coordinate system.

 We give at first
 our definition of
 the incoming--WIMP coordinate system
 as well as
 those of
 (the orientation of) the scattering plane
 and the (equivalent) recoil angle.
 Then
 we discuss
 the validation criterion
 in our Monte Carlo simulation
 by taking into account
 the cross section (nuclear form factor) suppression
 in detail.

\subsection{Definition of the incoming--WIMP coordinate system}
\label{sec:XYZ_chi}
\InsertSKPPlotS [10.5]
 {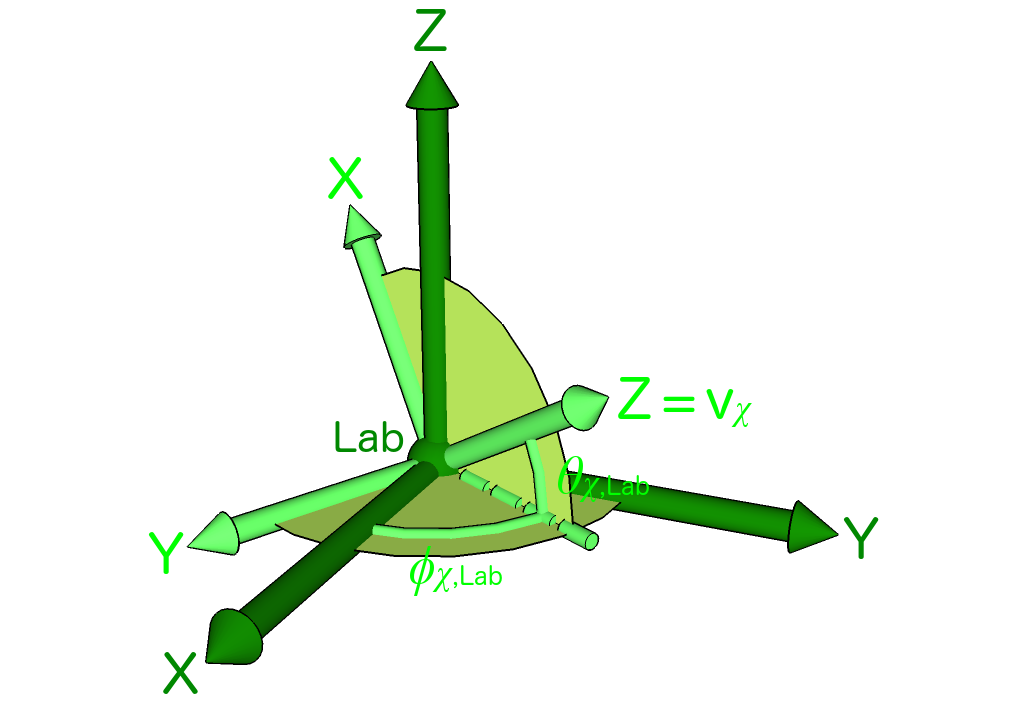}
 {The definition of
  the (light--green) incoming--WIMP coordinate system
  in the (dark--green) laboratory coordinate system.
  The $\zchi$--axis is defined as usual as
  the direction of the incident velocity of
  the incoming WIMP
  $\Vchi$.
  $\phichiLab$ and $\thetachiLab$ indicate
  the azimuthal angle and the elevation of
  the direction of
  $\Vchi$
  measured in the laboratory coordinate system,
  respectively.
  The $\xchi$--axis is perpendicular to the $\zchi$--axis
  and lies on the $\zLab$--$\zchi$ plane.
  Then
  the $\ychi$--axis is defined
  by the right--handed convention.%
  }

 In Fig.~\ref{fig:chi-Lab},
 we sketch the definition of
 the (light--green) incoming--WIMP coordinate system
 in the (dark--green) laboratory coordinate system%
\footnote{
 The transformation matrices
 between the incoming--WIMP
 and the laboratory coordinate systems
 will be given
 in Appendix \ref{appx:XYZ_chi}.
}.
 Note that,
 practically,
 the center of the incoming--WIMP coordinate system
 is at the position of the scattered target nucleus
 before scattering
 (see Fig.~\ref{fig:NR-chi-Lab}).
 The $\zchi$--axis is defined as usual as
 the direction of the incident velocity of
 the incoming WIMP
 $\Vchi$.
 $\phichiLab$ and $\thetachiLab$ indicate
 the azimuthal angle and the elevation of
 the direction of
 $\Vchi$
 measured in the laboratory coordinate system,
 respectively.
 The $\xchi$--axis is perpendicular to the $\zchi$--axis
 and lies on the $\zLab$--$\zchi$ plane.
 Then
 the $\ychi$--axis is defined
 by the right--handed convention.
 Note that
 the $\ychi$--axis lies always on the $\xLab$--$\yLab$ plane,
 since it is
 perpendicular to the $\xchi$--$\zLab$--$\zchi$ plane.

 Note also that,
 in our Monte Carlo simulation of
 3-D elastic WIMP--nucleus scattering events,
 the velocity (incident direction) of halo WIMPs
 in the laboratory
 and
 the Equatorial coordinate systems
 as well as
 the $\zchi$--axis of
 the incoming--WIMP coordinate system
 are {\em not fixed}
 as from the direction of the CYGNUS constellation.
 Interested readers can refer to Ref.~\cite{DMDDD-N}
 for the detailed discussions about
 (the annual and the diurnal modulations of)
 the anisotropy of
 the angular distributions of
 the 3-D WIMP velocity (flux)
 in the laboratory
 and the Equatorial coordinate systems.

\subsection{Generation of nuclear recoil directions}
\label{sec:phi_theta_NR_chi}
\InsertSKPPlotS [10.5]
 {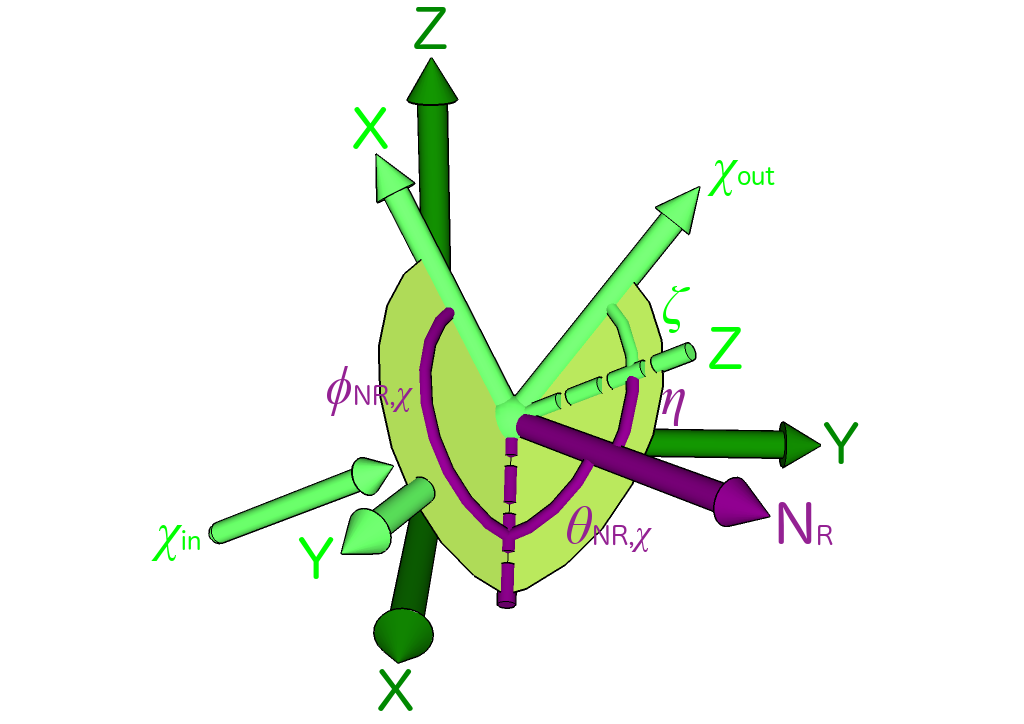}
 {A 3-D elastic WIMP--nucleus scattering event
  in the (light--green) incoming--WIMP
  and the (dark--green) laboratory coordinate systems.
  $\zeta$ and $\eta$ are
  the scattering angle of
  the outgoing WIMP $\chi_{\rm out}$
  and
  the recoil angle of
  the scattered target nucleus N$_{\rm R}$
  measured in the incoming--WIMP coordinate system
  of {\em this single} scattering event,
  respectively.
  While
  the azimuthal angle of
  the recoil direction of
  the scattered nucleus N$_{\rm R}$
  in this incoming--WIMP coordinate system,
  $\phiNRchi$,
  indicates the orientation of the scattering plane,
  the elevation of
  the recoil direction of N$_{\rm R}$,
  $\thetaNRchi$,
  is namely the complementary angle of
  the recoil angle
  $\eta$.%
  }

 At first,
 we sketch in Fig.~\ref{fig:NR-chi-Lab}
 the process of
 one single 3-D elastic WIMP--nucleus scattering event:
 $\chi_{\rm in/out}$ indicate
 the incoming and the outgoing WIMPs,
 respectively.
 While
 $\zeta$ indicates
 the scattering angle of
 the outgoing WIMP $\chi_{\rm out}$
 (measured from the $\zchi$--axis),
 $\eta$ is the recoil angle of
 the scattered target nucleus N$_{\rm R}$.

 It can be found firstly that,
 according to
 our definition of the incoming--WIMP coordinate system,
 the orientation of
 the ($\Vchiout$--$\zchi$--${\bf v}_{\rm N_R}$) scattering plane
 of {\em this single} scattering event
 (in the incoming--WIMP coordinate system)
 can be specified by
 the azimuthal angle of
 the recoil direction of
 the scattered nucleus,
 $\phiNRchi$,
 which
 should be azimuthal symmetric
 around the $\zchi$--axis
 and is thus
 generated with a constant probability
 in our simulation package:
\beq
     f_{{\rm N_R}, \chiin, \phi}(\phiNRchi)
  =  1
\~,
     ~~~~ ~~~~ ~~ 
     \phiNRchi \in (-\pi,~\pi]
\~.
\label{eqn:f_NR_phiNRchi}
\eeq
 Meanwhile,
 Fig.~\ref{fig:NR-chi-Lab} shows also that
 the elevation of
 the recoil direction of
 the scattered nucleus,
 $\thetaNRchi$,
 is namely the complementary angle of
 the recoil angle $\eta$:
\beq
      \thetaNRchi
  =   \frac{\pi}{2} - \eta
\~.
\label{eqn:thetaNRchi_eta}
\eeq
 Hence,
 in our simulation package,
 we use
\beq
      \thetaNRchi
 \in  [0,~\pi / 2]
\label{eqn:thetaNRchi_range}
\eeq
 as the ``equivalent'' recoil angle%
\footnote{
 Note that,
 without special remark,
 in this paper and our further works
 (e.g.~Refs.~\cite{DMDDD-NR, DMDDD-fv_eff}),
 we will use simply ``the recoil angle''
 to indicate ``the equivalent recoil angle $\thetaNRchi$''
 (not $\eta$).
}.

 In contrast to
 the simple constant generating probability
 $f_{{\rm N_R}, \chiin, \phi}(\phiNRchi)$
 given in Eq.~(\ref{eqn:f_NR_phiNRchi})
 for the orientation of the scattering plane
 $\phiNRchi$,
 the generating probability distribution of
 the recoil angle
 $\thetaNRchi$
 is more complicated and crucial.
 Below we discuss
 the cross section (nuclear form factor) suppression
 on the probability distribution of
 $\thetaNRchi$
 in detail%
\footnote{
 It would be important to emphasize here that,
 to the best of our knowledge,
 this should be the first time in literature that
 some constraints on
 the nuclear recoil angle/direction
 caused by
 (elastic) WIMP--nucleus scattering cross sections (nuclear form factors)
 have been considered
 in (3-D) WIMP scattering simulations.
}.
\subsubsection{Validation of 3-D elastic WIMP--nucleus scattering events}
\label{sec:dsigma_dthetaNRchi}

 For one WIMP event
 generated in the Galactic coordinate system
 and transformed step by step
 into the laboratory coordinate system
 with the velocity of
 $\Vchi(\vchiLab, \phichiLab, \thetachiLab)$,
 the kinetic energy
 can be given by
\beq
     \Echi
  =  \frac{1}{2} \mchi |\Vchi|^2
  =  \frac{1}{2} \mchi \vchiLab^2
\~.
\label{eqn:Echi}
\eeq
 Then
 the recoil energy of the scattered target nucleus
 in the incoming--WIMP coordinate system
 can be estimated by
 the recoil angle $\eta$
 or the equivalent recoil angle
 $\thetaNRchi$ as
\beq
     Q
  =  \bbrac{\frac{4 \mchi \mN}{(\mchi + \mN)^2} \~ \cos^2(\eta)}
     \Echi
  =  \bbrac{\afrac{2 \mrN^2}{\mN} \vchiLab^2}
     \sin^2(\thetaNRchi)
\~,
\label{eqn:QQ_thetaNRchi}
\eeq
 where
\beq
         \mrN
 \equiv  \frac{\mchi \mN}{\mchi + \mN}
\label{eqn:mrN}
\eeq
 is the reduced mass of
 the WIMP mass $\mchi$ and
 that of the target nucleus $\mN$.
 From Eq.~(\ref{eqn:QQ_thetaNRchi}),
 one can get that
\beq
     \Dd{Q}{\thetaNRchi}
  =  \bbrac{\afrac{2 \mrN^2}{\mN} \vchiLab^2}
     \sin(2 \thetaNRchi)
\~.
\label{eqn:dQQ_dthetaNRchi}
\eeq
 Hence,
 the differential cross section $d\sigma$
 given by
 the absolute value of the momentum transfer
 from the incident WIMP to the recoiling target nucleus,
\beq
     q
  =  |{\bf q}|
  =  \sqrt{2 \mN Q}
\~,
\label{eqn:qq}
\eeq
 can be obtained as
 \cite{SUSYDM96}
\beq
     d\sigma
  =  \frac{1}{\vchiLab^2}
     \afrac{\sigma_0}{4 \mrN^2} F^2(q) \~ dq^2
  =  \sigma_0 F^2(Q)
     \sin(2 \thetaNRchi) \~ d\thetaNRchi
\~.
\label{eqn:dsigma_thetaNRchi}
\eeq
 Then
 the differential WIMP--nucleus scattering cross section
 with respect to
 the recoil angle
 $\thetaNRchi$
 can generally be given by
\beq
     \Dd{\sigma}{\thetaNRchi}
  =  \bbigg{\sigmaSI \FSIQ + \sigmaSD \FSDQ}
     \sin(2 \thetaNRchi)
\~.
\label{eqn:dsigma_dthetaNRchi}
\eeq
 Here
 $\sigma_0^{\rm (SI, SD)}$ are
 the spin--independent (SI)/spin--dependent (SD) total cross sections
 ignoring the form factor suppression
 and
 $F_{\rm (SI, SD)}(Q)$ indicate the elastic nuclear form factors
 corresponding to the SI/SD WIMP interactions,
 respectively.
 Remind that
 the recoil energy $Q$
 is the function of the recoil angle $\thetaNRchi$
 given by Eq.~(\ref{eqn:QQ_thetaNRchi}).

 Finally,
 taking into account
 the proportionality of the WIMP flux
 to the incident velocity,
 the generating probability distribution of
 the recoil angle
 $\thetaNRchi$,
 which is proportional to
 the scattering event rate of
 incident halo WIMPs
 with an incoming velocity $\vchiLab$
 off target nuclei
 going into recoil angles of
 $\thetaNRchi \pm d\thetaNRchi / 2$
 with recoil energies of $Q \pm dQ / 2$,
 can generally be given by
\beqn
     f_{{\rm N_R}, \chiin, \theta}(\thetaNRchi)
 \=  \afrac{\vchiLab}{v_{\chi, {\rm cutoff}}}
     \aDd{\sigma}{\thetaNRchi}
     \non\\
 \=  \afrac{\vchiLab}{v_{\chi, {\rm cutoff}}}
     \bbigg{\sigmaSI \FSIQ + \sigmaSD \FSDQ}
     \sin(2 \thetaNRchi)
\~,
\label{eqn:f_NR_thetaNRchi}
\eeqn
 where
 $v_{\chi, {\rm cutoff}} \simeq 800$ km/s is
 a cut--off velocity of incident halo WIMPs
 in the laboratory coordinate system.

\subsubsection{WIMP--nucleus cross sections and nuclear form factors}
\label{sec:sigma_FQ_SI/SD}

 For the SI
 scalar WIMP interaction%
\footnote{
 Besides of the scalar interaction,
 WIMPs could also have a SI vector interaction
 with nuclei
 \cite{SUSYDM96}:
\beq
     \sigma_0^{\rm vector}
  =  \afrac{1}{64 \pi} \mrN^2
     \bBig{2 Z \brmp + (A - Z) \brmn}^2
\~,
\label{eqn:sigma0_vector}
\eeq
 where $b_{\rm (p, n)}$ are the effective
 vector couplings on protons and on neutrons,
 respectively.
 However,
 for Majorana WIMPs ($\chi = \Bar{\chi}$),
 e.g.~the lightest neutralino in supersymmetric models,
 there is no such vector interaction.
},
 the zero--momentum--transfer cross section
 in Eq.~(\ref{eqn:dsigma_dthetaNRchi})
 has been given by
 \cite{SUSYDM96}
\beq
     \sigmaSI
  =  \afrac{4}{\pi} \mrN^2
     \bBig{Z \frmp + (A - Z) \frmn}^2
  =  A^2 \afrac{\mrN}{\mrp}^2 \sigmapSI
\~.
\label{eqn:sigma0SI}
\eeq
 Here
 $\mrp$
 is the reduced mass of the WIMP mass $\mchi$
 and the proton mass $m_{\rm p}$,
 $Z$ is the atomic number of the target nucleus,
 i.e.~the number of protons,
 $A$ is the atomic mass number,
 $A - Z$ is then the number of neutrons,
 $f_{\rm (p, n)}$ are the effective scalar couplings of WIMPs
 on protons p and on neutrons n,
 respectively,
 and
\beq
     \sigmapSI
  =  \afrac{4}{\pi} \mrp^2 |\frmp|^2
\label{eqn:sigmapSI}
\eeq
 is the SI scalar WIMP--nucleon cross section.
 The theoretical prediction
 for the lightest supersymmetric neutralino:
 the scalar couplings are approximately the same
 on protons and on neutrons,
\(
         \frmn
 \simeq  \frmp
\),
 has been adopted here
 and
 the tiny mass difference between a proton and a neutron
 has been neglected.

 On the other hand,
 the SD axial--vector WIMP--nucleus cross section
 in Eq.~(\ref{eqn:dsigma_dthetaNRchi})
 can be expressed as
 \cite{SUSYDM96}
\beqn
     \sigmaSD
 \=  \afrac{32}{\pi} G_F^2 \~ \mrN^2
     \afrac{J + 1}{J} \bBig{\Srmp \armp + \Srmn \armn}^2
     \non\\
 \=  \frac{4}{3} \afrac{J + 1}{J} \afrac{\mrN}{\mrp}^2
     \bbrac{\Srmp + \Srmn \afrac{\armn}{\armp}}^2
     \sigmapSD
\~.
\label{eqn:sigma0SD}
\eeqn
 Here $G_F$ is the Fermi constant,
 $J$ is the total spin of the target nucleus,
 $\expv{S_{\rm (p, n)}}$ are the expectation values of
 the proton and neutron group spins
 (see Table \ref{tab:Sp/n}
  for the list of the default spin values of the nuclei
  used in our simulation package),
 $a_{\rm (p, n)}$ are the effective SD axial--vector WIMP couplings
 on protons and on neutrons,
 respectively,
 and
 the SD WIMP cross section on protons or on neutrons
 can be given by
\beq
     \sigma_{\chi {\rm (p, n)}}^{\rm SD}
  =  \afrac{24}{\pi} G_F^2 \~ m_{\rm r, (p, n)}^2 |a_{\rm (p, n)}|^2
\~.
\label{eqn:sigmap/nSD}
\eeq
\begin{table}[t!]
\small
\begin{center}
\renewcommand{\arraystretch}{1.85}
\begin{tabular}{|| c   c   c   c   c   c ||}
\hline
\hline
 \makebox[2  cm][c]{Isotope}        &
 \makebox[1.5cm][c]{$Z$}            & \makebox[1.5cm][c]{$J$}     &
 \makebox[2.5cm][c]{$\Srmp$}        & \makebox[2.5cm][c]{$\Srmn$} &
 \makebox[4.5cm][c]{Natural abundance (\%)} \\
\hline
\hline
 $\rmXA{Li}{7}$   &  3 & 3/2 &                    0.497   &                    0.004   &
  92.41  \\
\hline
 $\rmXA{O}{17}$   &  8 & 5/2 &                    0       &                    0.495  &
   0.038 \\
\hline
 $\rmXA{F}{19}$   &  9 & 1/2 &                    0.441   &  \hspace{-1.8ex}$-$0.109   &
 100     \\
\hline
 $\rmXA{Na}{23}$  & 11 & 3/2 &                    0.248   &                    0.020   &
 100     \\
\hline
 $\rmXA{Al}{27}$  & 13 & 5/2 &                    0.343   &                    0.030   &
 100     \\
\hline
 $\rmXA{Si}{29}$  & 14 & 1/2 &  \hspace{-1.8ex}$-$0.002   &                    0.130   &
   4.68  \\
\hline
 $\rmXA{Cl}{35}$  & 17 & 3/2 &  \hspace{-1.8ex}$-$0.059   &  \hspace{-1.8ex}$-$0.011   &
  75.78  \\
\hline
 $\rmXA{Cl}{37}$  & 17 & 3/2 &  \hspace{-1.8ex}$-$0.058   &                    0.050   &
  24.22  \\
\hline
 $\rmXA{K}{39}$   & 19 & 3/2 &  \hspace{-1.8ex}$-$0.180   &                    0.050   &
  93.26  \\
\hline
 $\rmXA{Ge}{73}$  & 32 & 9/2 &                    0.030   &                    0.378   &
   7.73  \\
\hline
 $\rmXA{Nb}{93}$  & 41 & 9/2 &                    0.460   &                    0.080   &
 100     \\
\hline
 $\rmXA{Te}{125}$ & 52 & 1/2 &                    0.001   &                    0.287   &
   7.07  \\
\hline
 $\rmXA{I}{127}$  & 53 & 5/2 &                    0.309   &                    0.075   &
 100     \\
\hline
 $\rmXA{Xe}{129}$ & 54 & 1/2 &                    0.028   &                    0.359   &
  26.44  \\
\hline
 $\rmXA{Xe}{131}$ & 54 & 3/2 &  \hspace{-1.8ex}$-$0.009   &  \hspace{-1.8ex}$-$0.227   &
  21.18  \\
\hline
 $\rmXA{Cs}{133}$ & 55 & 7/2 &  \hspace{-1.8ex}$-$0.370   &                    0.003   &
 100     \\
\hline
 $\rmXA{W}{183}$  & 74 & 1/2 &                    0       &  \hspace{-1.8ex}$-$0.031   &
  14.31  \\
\hline
\hline
\end{tabular}
\end{center}
\caption{
 List of the default spin values of the nuclei
 used in our simulation package
 \cite{Engel95, Ressell97,
       Tovey00, Giuliani05,
       WebElements}.
}
\label{tab:Sp/n}
\end{table}

 Substituting Eqs.~(\ref{eqn:sigma0SI}) and (\ref{eqn:sigma0SD})
 into Eq.~(\ref{eqn:f_NR_thetaNRchi}),
 the validation criterion of
 our 3-D elastic WIMP--nucleus scattering simulation
 can be expressed as
\beqn
\conti
     f_{{\rm N_R}, \chiin, \theta}(\thetaNRchi)
     \non\\
 \=  \afrac{\vchiLab}{v_{\chi, {\rm cutoff}}}
     \afrac{\mrN}{\mrp}^2
     \non\\
 \conti ~~~~ \times
     \cBigg{  A^2
              \sigmapSI \FSIQ
            + \frac{4}{3} \afrac{J + 1}{J} \bbrac{\Srmp + \Srmn \afrac{\armn}{\armp}}^2
              \sigmapSD \FSDQ  }
     \sin(2 \thetaNRchi)
\~.
\label{eqn:f_NR_thetaNRchi-SISD}
\eeqn
 Additionally,
 as default setup
 in our simulation package,
 we adopt
 the commonly used analytic form
 for the elastic nuclear form factor
 \cite{SUSYDM96}
\beq
     F_{\rm SI}^2(Q)
  =  \bfrac{3 j_1(q R_1)}{q R_1}^2 e^{-(q s)^2}
\~,
\label{eqn:FQ_SI_WS}
\eeq
 as well as
 the thin--shell form factor
 \cite{Lewin96}
\beq
     F_{\rm SD}^2(Q)
  =  \cleft{\renewcommand{\arraystretch}{1.5}
            \begin{array}{l l l}
             j_0^2(q R_1)                      \~, & ~~~~ ~~~~ & 
             {\rm for}~q R_1 \le 2.55~{\rm or}~q R_1 \ge 4.5 \~, \\
             {\rm const.} \simeq 0.047         \~, &           &
             {\rm for}~2.5 5 \le q R_1 \le 4.5 \~,
            \end{array}}
\label{eqn:FQ_SD_TS}
\eeq
 for the SI and SD WIMP--nucleus cross sections,
 respectively.
 Here
 $j_1(x)$ and $j_0(x)$ are the spherical Bessel functions,
 for the effective nuclear radius we use
\beq
     R_1
  =  \sqrt{R_A^2 - 5 s^2}
\~,
\label{eqn:R1}
\eeq
 with
\beq
     R_A
 \simeq
     1.2 \~ A^{1 / 3} \~ {\rm fm}
\~,
\label{eqn:RA}
\eeq
 and a nuclear skin thickness
\beq
     s
 \simeq
     1 \~ {\rm fm}
\~.
\label{eqn:ss}
\eeq
\begin{figure} [t!]
\begin{center}
 \includegraphics [width = 12 cm] {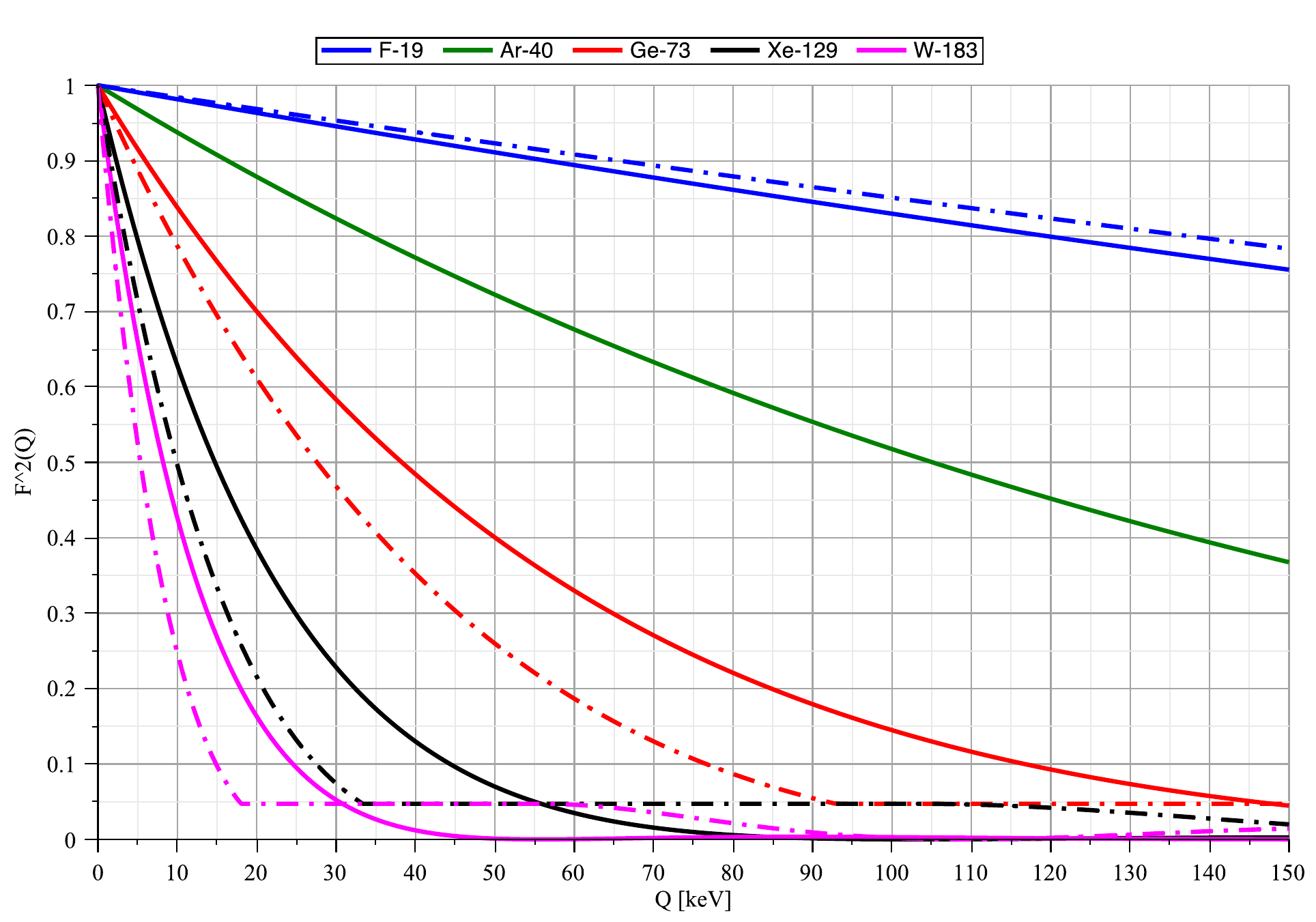}
\end{center}
\caption{
 Nuclear form factors of
 the $\rmF$     (blue),
 the $\rmAr$    (green),
 the $\rmGe$    (red),
 the $\rmXe$    (black),
 and the $\rmW$ (magenta) nuclei
 as functions of the recoil energy.
 The solid and dash--dotted curves
 indicate the form factors corresponding to
 the SI and SD cross sections,
 $F_{\rm SI}^2(Q)$ and $F_{\rm SD}^2(Q)$,
 given in Eqs.~(\ref{eqn:FQ_SI_WS}) and (\ref{eqn:FQ_SD_TS}),
 respectively.
}
\label{fig:FQ}
\end{figure}

 In Fig.~\ref{fig:FQ},
 we show
 the recoil--energy dependences of
 the nuclear form factors corresponding to
 the SI (solid) and SD (dash--dotted) cross sections,
 $F_{\rm SI}^2(Q)$ and $F_{\rm SD}^2(Q)$,
 given in Eqs.~(\ref{eqn:FQ_SI_WS}) and (\ref{eqn:FQ_SD_TS}),
 respectively.
 Five frequently used target nuclei
 have been considered:
 $\rmF$     (blue),
 $\rmAr$    (green),
 $\rmGe$    (red),
 $\rmXe$    (black),
 and $\rmW$ (magenta).
 The sharply enlarged nuclear form factor suppression
 in the validation criterion
 (\ref{eqn:f_NR_thetaNRchi}) or (\ref{eqn:f_NR_thetaNRchi-SISD})
 with the increased mass of the target nucleus
 can be seen clearly.

 In Refs.~\cite{DMDDD-NR} and \cite{DMDDD-fv_eff},
 we will discuss in detail
 the effects of the cross section (nuclear form factor) suppression
 on the angular distributions of
 the nuclear recoil direction (flux)/energy
 as well as
 on the 3-D {\em effective} velocity distribution of
 incident halo WIMPs
 separately.

\subsubsection{WIMP--mass dependence of the recoil energy}
\label{sec:Qmax_rms-mchi}
\begin{figure} [t!]
\begin{center}
 \includegraphics [width = 12 cm] {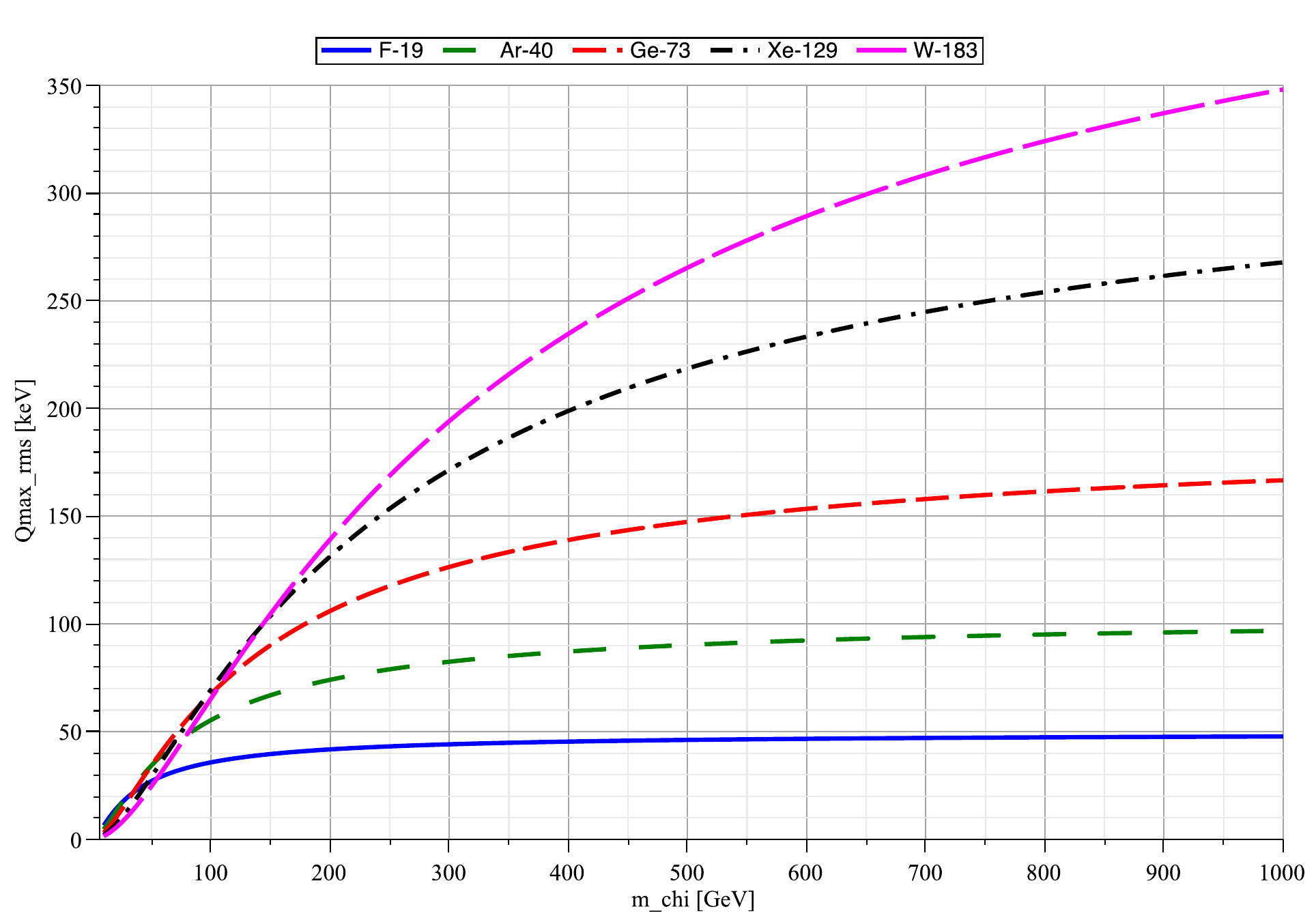}
\end{center}
\caption{
 The WIMP--mass dependence of
 the maximum of the recoil energy,
 $Q_{\rm max, rms}$,
 given by Eq.~(\ref{eqn:Qmax_rms}).
 Five frequently used target nuclei:
 $\rmF$     (solid        blue),
 $\rmAr$    (rare--dashed green),
 $\rmGe$    (dashed       red),
 $\rmXe$    (dash--dotted black),
 and $\rmW$ (long--dashed magenta)
 have been considered.
}
\label{fig:Qmax_rms-mchi}
\end{figure}

 For readers' reference,
 in Fig.~\ref{fig:Qmax_rms-mchi}
 we show
 the WIMP--mass dependence of
 the maximum (prefactor) of the recoil energy $Q$
 given by Eq.~(\ref{eqn:QQ_thetaNRchi})
 with the monotonic root--mean--square velocity
 of incident halo WIMPs:
\beq
     Q_{\rm max, rms}
  =  \afrac{2 \mrN^2}{\mN} v_{\rm rms, Lab}^2
\~,
\label{eqn:Qmax_rms}
\eeq
 for five frequently used target nuclei:
 $\rmF$     (solid        blue),
 $\rmAr$    (rare--dashed green),
 $\rmGe$    (dashed       red),
 $\rmXe$    (dash--dotted black),
 and $\rmW$ (long--dashed magenta).
 Here,
 although
 it would be somehow inconsistent
 with our observations presented
 in Ref.~\cite{DMDDD-N}
 (see detailed discussions therein),
 we use
 the shifted Maxwellian velocity distribution function
 \cite{SUSYDM96}:
\beq
     f_{1, \sh}(\vchiLab)
  =  \frac{1}{\sqrt{\pi}} \afrac{\vchiLab}{v_0 \ve}
     \bbigg{  e^{-(\vchiLab - \ve)^2 / v_0^2}
            - e^{-(\vchiLab + \ve)^2 / v_0^2}  }
\~,
\label{eqn:f1v_sh}
\eeq
 as a useful approximation to
 the radial component (magnitude) of
 the 3-D WIMP velocity distribution
 in the Equatorial/laboratory coordinate systems,
 where
 $\ve$ is the time--dependent
 Earth's velocity in the Galactic frame
 \cite{Freese88,
       SUSYDM96}:
\beq
     \ve(t)
  =  v_0 \bbrac{1.05 + 0.07 \cos\afrac{2 \pi (t - \tp)}{1~{\rm yr}}}
\~,
\label{eqn:ve}
\eeq
 with $\tp \simeq$ June 2nd,
 the date
 on which
 the Earth's orbital speed
 is maximal.
 Then
 the root--mean--square velocity of
 incident halo WIMPs
 can be obtained as
\beqn
     v_{\rm rms, Lab}^2
 \=  \Expv{\vchiLab^2}_{\sh}
  =  \intz \vchiLab^2 \~ f_{1, \sh}(\vchiLab) \~ d\vchiLab
  =  \afrac{3}{2} v_0^2 + \ve^2
     \non\\
 \eqnsimeq  (355~{\rm km/s})^2
\~.
\label{eqn:expv_v2_sh}
\eeqn
 In the last line,
 the time dependence of $\ve(t)$
 has been ignored and
 $\ve = 1.05 \~ v_0$
 is used.

 In Refs.~\cite{DMDDD-NR} and \cite{DMDDD-fv_eff},
 we will demonstrate
 the WIMP--mass dependence of
 the cross section (nuclear form factor) suppression
 on the angular recoil--direction/energy distributions
 as well as
 on the 3-D WIMP effective velocity distribution
 separately.

\section{Summary}

 As the preparation of
 our development of
 data analysis procedures
 for using and/or combining 3-dimensional information
 offered by directional direct Dark Matter detection experiments
 in the future,
 we finally achieved
 our double--Monte Carlo scattering--by--scattering simulation of
 3-dimensional elastic WIMP--nucleus scattering process,
 which can provide
 3-D information
 (the magnitude,
  the direction,
  and the incoming/scattering time) of
 each incident halo WIMP
 as well as
 the experimentally measurable
 recoil direction
 and recoil energy of
 the WIMP--scattered target nucleus
 event by event
 in different celestial coordinate systems.

 In this paper,
 we described at first
 the overall workflow of our simulation procedure.
 After the summary of
 the MC generation process of
 the 3-D velocity information of
 Galactic WIMPs
 and
 the transformations of
 the 3-D (velocity) information
 between different celestial coordinate systems,
 we introduced the incoming--WIMP coordinate system
 for describing the 3-D WIMP--nucleus scattering process
 and derived the validation criterion
 in our MC generation of
 the 3-D recoil information of
 scattered target nuclei,
 which is basically according to
 the cross section (nuclear form factor) suppression
 on the {\em recoil--angle--dependent} recoil energy.

 Currently,
 several approximations
 about the Earth's orbital motion
 in the Solar system
 and the observation periods/daily shifts
 have been used in our simulation package.
 First,
 the Earth's orbit around the Sun is perfectly circular
 on the Ecliptic plane
 and the orbital speed is thus a constant.
 Second,
 the date of the vernal equinox is exactly fixed
 at the end of the May 20th (the 79th day) of
 a 365-day year
 and the few extra hours
 in an actual Solar year
 have been neglected.
 Nevertheless,
 considering the very low WIMP scattering event rate
 and thus maximal a few (tens) of total WIMP events
 observed in at least a few tens (or even hundreds) of days
 (an optimistic overall event rate of $\lsim \~ {\cal O}(1)$ event/day)
 for the first--phase analyses,
 these approximations should be acceptable.

 Hopefully,
 this (and more works fulfilled in the future)
 could help our colleagues
 to develop analysis methods
 for understanding
 the astrophysical and particle properties of
 Galactic WIMPs
 as well as
 the structure of Dark Matter halo
 by using directional direct detection data.

\subsubsection*{Acknowledgments}

 The author appreciates
 N.~Bozorgnia and P.~Gondolo
 for useful discussions
 about the transformations between the celestial coordinate systems.
 The author would like to thank
 the friendly hospitality of
 the Gran Sasso Science Institute
 as well as
 the pleasant atmosphere of
 the W101 Ward and the Cancer Center of
 the Kaohsiung Veterans General Hospital,
 where part of this work was completed.
 This work
 was strongly encouraged by
 the ``{\it Researchers working on
 e.g.~exploring the Universe or landing on the Moon
 should not stay here but go abroad.}'' speech.

\appendix
\setcounter{equation}{0}
\setcounter{figure}  {0}
\setcounter{table}   {0}
\renewcommand{\theequation}{A\arabic{equation}}
\renewcommand{\thefigure}  {A\arabic{figure}}
\renewcommand{\thetable}   {A\arabic{table}}
%

%
%
%
 \renewcommand{\arraystretch}{1.5}
\section{Definitions of and transformations between our coordinate systems}
\label{appx:XYZ}

 In this section,
 we review briefly our definitions of
 the laboratory--independent
 (Galactic,
  Ecliptic,
  Equatorial,
  and Earth)
 coordinate systems
 as well as
 the laboratory--dependent
 (horizontal and laboratory)
 coordinate systems
 used in our simulation package.
 We also summarize
 the matrices
 needed for the transformations (of
 the 3-D velocity)
 between these coordinate systems.

 Discussions about our coordinate systems
 as well as
 the detailed derivations of the transformation matrices
 between them
 can be found in Ref.~\cite{DMDDD-N}.

\subsection{Laboratory--independent coordinate systems}
\label{appx:XYZ_G-S-Eq}

 We consider
 the Galactic,
 the Ecliptic,
 and the Equatorial coordinate systems
 at first.

\subsubsection{Definitions}
\label{appx:def-XYZ_G-S-Eq}

 In Fig.~\ref{fig:v_Sun_Eq-S-G-rotated},
 we show
 the definitions of
 and the relative orientations between
 the (black) Galactic,
 the (red) Ecliptic,
 and the (blue) Equatorial coordinate systems
 (on the date of the vernal equinox).

\InsertSKPPlotS
 {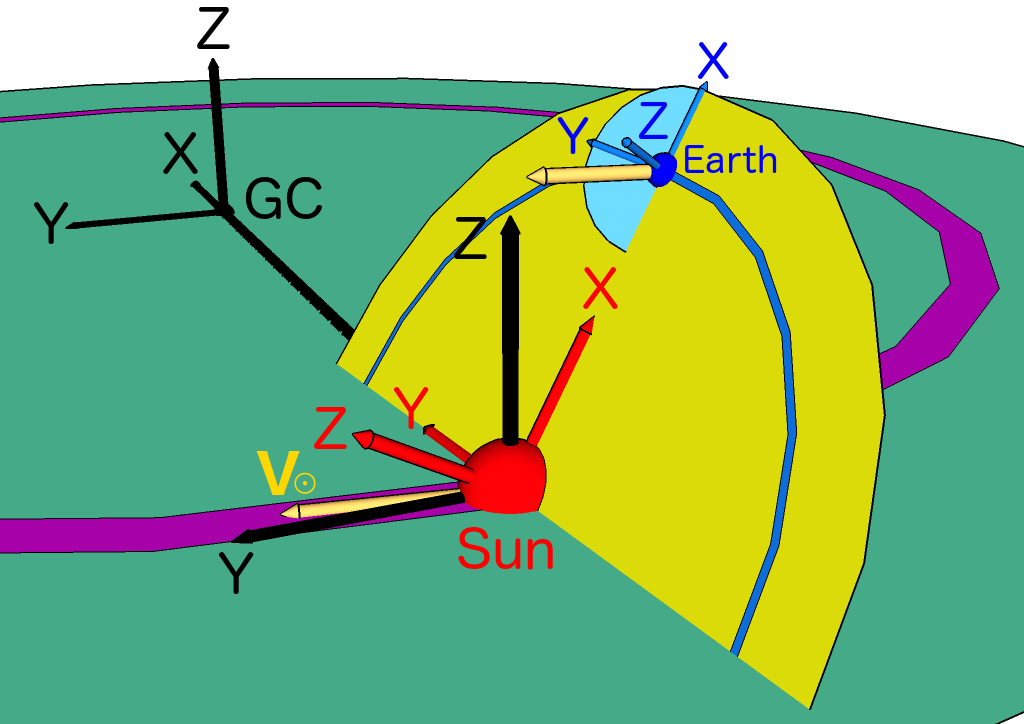}
 {The definitions of
  and the relative orientations between
  the (black) Galactic,
  the (red) Ecliptic,
  and the (blue) Equatorial coordinate systems
  (on the date of the vernal equinox).
  While
  the magenta circular band
  indicates an approximate path of
  the orbital motion of the Solar system
  in the Galaxy
  and
  the blue circular band
  the Earth's orbit around the Sun,
  the additional (golden) arrows
  indicate
  the direction of the movement of the Solar system
  around the Galactic center.
  See the text for the detailed descriptions.%
  }

 Firstly,
 the origin of the Galactic coordinate system
 is at the (approximate) Galactic Center (GC).
 The primary direction (the $\xG$--axis)
 points from the Solar center to GC
 and
 the $\zG$--axis
 to the Galactic North Pole (GNP).
 Then
 the right--handed convention
 is used for defining the $\yG$--axis
 and the fundamental ($\xG - \yG$) plane
 is the approximate Galactic plane
 \cite{Wiki-Galactic}.

 Meanwhile,
 the origins of
 the Ecliptic
 and the Equatorial coordinate systems
 are at
 the center of the Sun
 and that of the Earth,
 respectively.
 The common primary direction (the $\xS$/$\xEq$--axis)
 is the direction
 pointing from the Solar center to that of the Earth
 at 12 midnight (the {\em end}) of
 the date of the vernal equinox,
 the $\zS$-- and the $\zEq$--axes
 are perpendicular to the (yellow) Ecliptic
 and the (blue) Equatorial planes,
 respectively,
 and their $\yS$-- and $\yEq$--axes
 are then defined as usual by
 the right--handed convention.

 Additionally,
 in Fig.~\ref{fig:v_Sun_Eq-S-G-rotated},
 we also draw
 two (golden) arrows
 to indicate
 the direction of the movement of the Solar system
 towards the CYGNUS constellation.
 Note that
 the moving direction of the Solar system
 is not parallel to,
 but only approximately along
 the $\yG$--axis,
 (with an included angle of 8.87$^{\circ}$ = 35.48$^{\rm m}$),
 nor on the (approximate) Galactic plane
 (0.60$^{\circ}$ above).

 Note also that
 the Ecliptic coordinate system
 only moves approximately linearly
 with the Solar Galactic orbital velocity $\vSunG = |\VSunG| \simeq 220$ km/s
 and
 the tiny Galactic orbital rotation
 of the Solar system
 is considered to be imperceptible,
 whereas
 the Equatorial coordinate system
 moves orbitally around
 (and also linearly with) the Sun,
 but doesn't rotate.
 These mean that
 the axes of
 the Galactic,
 the Ecliptic,
 and the Equatorial coordinate systems
 defined in our simulation package
 are all {\em fixed}
 (see Table \ref{tab:CSs}
  for the summary of
  the styles of the movements and the rotations
  of different celestial coordinate systems).

\subsubsection{Transformation matrices}
\label{appx:Ma-XYZ_G-S-Eq}

 At first,
 by definition,
 the transformation matrix
 from the Ecliptic coordinate system
 to the Equatorial coordinate system
 can be given directly as
\cheqnXa{A}
\beqn
     \MaSEq
  =  \left[\begin{array}{c c c}
             1 ~&~   0              ~&~  0              \\
             0 ~&~  \cos(\psiEarth) ~&~ \sin(\psiEarth) \\
             0 ~&~ -\sin(\psiEarth) ~&~ \cos(\psiEarth) \\
           \end{array}\right]
  =  \left[\begin{array}{c c c}
             1 ~&~  0       ~&~ 0       \\
             0 ~&~  0.91775 ~&~ 0.39715 \\
             0 ~&~ -0.39715 ~&~ 0.91775 \\
           \end{array}\right]
\~,
\label{eqn:Ma_S_Eq}
\eeqn
 and,
 conversely,
\cheqnXb{A}
\beq
     \MaEqS
  =  \left[\begin{array}{c c c}
             1 ~&~  0              ~&~   0              \\
             0 ~&~ \cos(\psiEarth) ~&~ -\sin(\psiEarth) \\
             0 ~&~ \sin(\psiEarth) ~&~  \cos(\psiEarth) \\
           \end{array}\right]
  =  \left[\begin{array}{c c c}
             1 ~&~ 0       ~&~  0       \\
             0 ~&~ 0.91775 ~&~ -0.39715 \\
             0 ~&~ 0.39715 ~&~  0.91775 \\
           \end{array}\right]
\~,
\label{eqn:Ma_Eq_S}
\eeq
\cheqnX{A}
  where $\psiEarth = 23.4^{\circ}$
  is the Earth's obliquity.

 On the other hand,
 the directions of the Galactic Center
 and the Galactic North Pole
 in the Equatorial coordinate system
 can be expressed by
\cheqnXa{A}
\beqn
     \xGEq
 \=  \left[\begin{array}{c c c}
            \cos(\thetaGC) \cos(\phiGC) ~&~
            \cos(\thetaGC) \sin(\phiGC) ~&~
            \sin(\thetaGC)               \\
           \end{array}\right]_{\rm Eq}
     \non\\
 \eqncong
     \left[\begin{array}{c c c}
             0.05495 ~&~
             0.87340 ~&~
            -0.48389  \\
           \end{array}\right]_{\rm Eq}
\~,
\label{eqn:GC_Eq}
\eeqn
 and
\cheqnXb{A}
\beqn
     \zGEq
 \=  \left[\begin{array}{c c c}
            \cos(\thetaGNP) \cos(\phiGNP) ~&~
            \cos(\thetaGNP) \sin(\phiGNP) ~&~
            \sin(\thetaGNP)                \\
           \end{array}\right]_{\rm Eq}
     \non\\
 \eqncong
      \left[\begin{array}{c c c}
            0.86769 ~&~
            0.19793 ~&~
            0.45601  \\
           \end{array}\right]_{\rm Eq}
\~,
\label{eqn:GNP_Eq}
\eeqn
\cheqnX{A}%
 respectively,
 where
 we have adopted the values
 provided by Ref.~\cite{Wiki-Galactic}%
\footnote{
 Note that
 the common $\xS$/$\xEq$--axis
 defined in our Ecliptic and Equatorial coordinate systems
 points from the center of the Sun to that of the Earth
 and is thus opposite to
 the conventional astronomical definition.
 Hence,
 the right ascensions of GC
 and GNP
 in the Equatorial coordinate system
 given here
 differ from the values given in Ref.~\cite{Wiki-Galactic}
 by $12^{\rm h} = 180^{\circ}$.
\label{fn:phi_GC_Eq}
}
\beq
     \phiGC
  =  5^{\rm h} 45.6^{\rm m}
  =  86.40^{\circ}
\~,
     ~~~~ ~~~~ ~~~~ ~~~~ ~~~~ 
     \thetaGC
  = -28.94^{\circ}
\~,
\label{eqn:ang_GC_Eq}
\eeq
 and
\beq
     \phiGNP
  =  51.4^{\rm m}
  =  12.85^{\circ}
\~,
     ~~~~ ~~~~ ~~~~ ~~~~ ~~~~ ~ 
     \thetaGNP
  =  27.13^{\circ}
\~,
\label{eqn:ang_GNP_Eq}
\eeq
 as the right ascensions and the declinations of
 GC
 and GNP
 in the Equatorial coordinate system,
 respectively.

 Then,
 by combining Eqs.~(\ref{eqn:GC_Eq}) and (\ref{eqn:GNP_Eq}),
 the $\yG$--axis of the Galactic coordinate system
 in the Equatorial coordinate system
 can be calculated by
\cheqnXNx{A}{-3}{c}
\beqn
     \yGEq
 \=  \zGEq \times \xGEq
     \non\\
 \=  \footnotesize
     \left[\renewcommand{\arraystretch}{1.75}
           \begin{array}{c}
              \cos(\thetaGNP) \sin(\phiGNP)
              \sin(\thetaGC)
            - \sin(\thetaGNP)
              \cos(\thetaGC)  \sin(\phiGC)  \\
              \sin(\thetaGNP)
              \cos(\thetaGC)  \cos(\phiGC)
            - \cos(\thetaGNP) \cos(\phiGNP)
              \sin(\thetaGC)                \\
              \cos(\thetaGNP) \cos(\thetaGC)
              \bBig{\cos(\phiGNP) \sin(\phiGC) - \sin(\phiGNP) \cos(\phiGC)} \\
           \end{array}\right]_{\rm Eq}^{\rm T}
     \non\\
 \eqncong
      \left[\begin{array}{c c c}
            -0.49406 ~&~
             0.44492 ~&~
             0.74696  \\
           \end{array}\right]_{\rm Eq}
\~.
\label{eqn:Y_G_Eq}
\eeqn
\cheqnXN{A}{2}%
 Hence,
 the transformation matrices
 between the Equatorial
 and the Galactic coordinate systems
 can be given by
\cheqnXa{A}
\beq
     \MaEqG
  =  \left[\begin{array}{c}
            \xGEq \\
            \yGEq \\
            \zGEq \\
           \end{array}\right]
  =  \left[\begin{array}{c c c}
             0.05495 ~&~ 0.87340 ~&~ -0.48389 \\
            -0.49406 ~&~ 0.44492 ~&~  0.74696 \\
             0.86769 ~&~ 0.19793 ~&~  0.45601 \\
           \end{array}\right]
\~,
\label{eqn:Ma_Eq_G}
\eeq
 and,
 conversely,
\cheqnXb{A}
\beq
     \MaGEq
  =  \left[\begin{array}{c c c}
             0.05495 ~&~ -0.49406 ~&~ 0.86769 \\
             0.87340 ~&~  0.44492 ~&~ 0.19793 \\
            -0.48389 ~&~  0.74696 ~&~ 0.45601 \\
           \end{array}\right]
\~.
\label{eqn:Ma_G_Eq}
\cheqnX{A}
\eeq

 Moreover,
 by combining the transformation matrices
 between the Equatorial
 and the Ecliptic coordinate systems
 in Eqs.~(\ref{eqn:Ma_Eq_S}) and (\ref{eqn:Ma_S_Eq}),
 the transformations
 between the Ecliptic
 and the Galactic coordinate systems
 can be obtained by
\cheqnXa{A}
\beq
     \MaGS
  =  \MaEqS \MaGEq
  =  \left[\begin{array}{c c c}
             0.05495 ~&~ -0.49406 ~&~ 0.86769 \\
             0.99374 ~&~  0.11168 ~&~ 0.00055 \\
            -0.09723 ~&~  0.86223 ~&~ 0.49711 \\
           \end{array}\right]
\~,
\label{eqn:Ma_G_S}
\eeq
 and
\cheqnXb{A}
\beq
     \MaSG
  =  \MaEqG \MaSEq
  =  \left[\begin{array}{c c c}
             0.05495 ~&~ 0.99374 ~&~ -0.09723 \\
            -0.49406 ~&~ 0.11168 ~&~  0.86223 \\
             0.86769 ~&~ 0.00055 ~&~ 0.49711  \\
           \end{array}\right]
\~.
\label{eqn:Ma_S_G}
\eeq
\cheqnX{A}
\subsubsection{Direction of the Galactic movement of the Solar system}
\label{appx:v_Sun_G}

 For readers' reference,
 the direction
 (the right ascension and the declination) of
 the Galactic movement of the Solar system
 towards the CYGNUS constellation
 in the Galactic,
 the Ecliptic,
 and the Equatorial coordinate systems
 are summarized here%
\footnote{
 Note that,
 as reminded in footnote \ref{fn:phi_GC_Eq},
 the right ascensions of the CYGNUS constellation
 in the Equatorial coordinate system
 given here
 differ from the values given in Ref.~\cite{Bandyopadhyay10}
 by $12^{\rm h} = 180^{\circ}$.
}:
\beq
     \phiCYGNUS[G]
  =   5.41^{\rm h}
  =  81.22^{\circ}
\~,
     ~~~~ ~~~~ ~~~~ ~~~~ ~~~~ ~~\~ 
     \thetaCYGNUS[G]
  =  0.60^{\circ}
\~,
\label{eqn:and_CYGNUS_G}
\eeq
\beq
     \phiCYGNUS[S]
  =   10.06^{\rm h}
  =  150.90^{\circ}
\~,
     ~~~~ ~~~~ ~~~~ ~~~~ ~~~~ 
     \thetaCYGNUS[S]
  =  57.40^{\circ}
\~,
\label{eqn:ang_CYGNUS_S}
\eeq
 and \cite{Bandyopadhyay10}
\beq
     \phiCYGNUS
  =    8.62^{\rm h}
  =  129.30^{\circ}
\~,
     ~~~~ ~~~~ ~~~~ ~~~~ ~~~~ \:\! 
     \thetaCYGNUS
  =  42^{\circ}
\~.
\label{eqn:ang_CYGNUS_Eq}
\eeq
 Detailed derivations can be found
 in Ref.~\cite{DMDDD-N}.

\subsection{Earth coordinate system}
\label{appx:XYZ_E}

 As the connection
 of the Equatorial and Ecliptic coordinate systems
 to the horizontal and laboratory coordinate systems
 \cite{Bandyopadhyay10},
 we defined the Earth coordinate system
 in our simulation package.

\subsubsection{Definition}
\label{appx:def-XYZ_E}
\InsertSKPPlotS
 {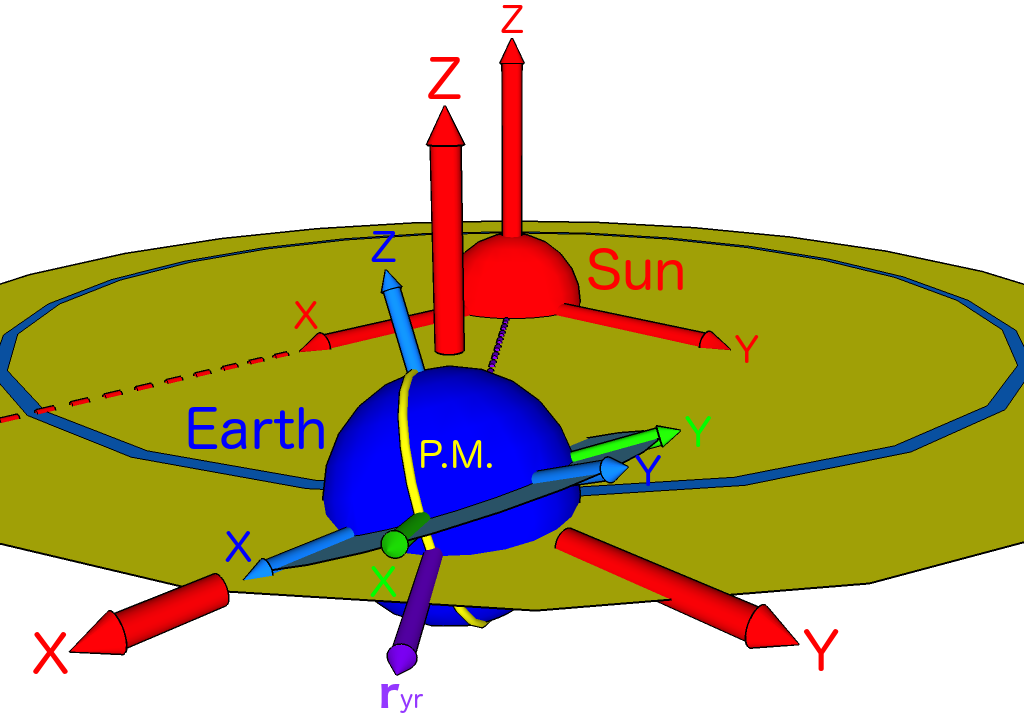}
 {The definition of
  the (light--green) Earth coordinate system
  at 12 midnight (the beginning)
  (i.e.,
   when the (yellow) Prime Meridian
   (the longitude 0$^{\circ}$)
   passes the purple arrow
   pointing from the Solar center to that of the Earth)
  of each single ``Solar'' day.
  The (red) Ecliptic and the (blue) Equatorial coordinate systems
  as well as
  the (blue) Earth's orbit around the Sun
  are also given here.
  See the text for the detailed description.%
  }

 As shown in Fig.~\ref{fig:E-Eq-S},
 we define
 the Earth coordinate system
 as follows:
 while
 the origin is also
 located at the Earth's center
 and
 the $\zE$--axis is still
 the Earth's north polar axis,
 the primary direction (the $\xE$--axis)
 points now from the Earth's center
 to the Prime Meridian
 (the longitude 0$^{\circ}$)
 at 12 midnight (the {\em beginning})
 (i.e.,
  when the Prime Meridian
  passes the direction
  pointing from the Solar center to that of the Earth)
 of each single ``Solar'' day.
 The fundamental ($\xE - \yE$) plane is again
 the Equatorial plane
 and the right--hand convention is used
 to define the $\yE$--axis.

 Note that,
 for each single (Solar) day,
 the Earth coordinate system
 is fixed with the direction of the Prime Meridian
 at (UTC) 12 midnight,
 but {\em rotates} with the Earth
 during the its orbital motion around the Sun.
 This means that
 our Earth coordinate system
 changes daily and discretely
 (see Table \ref{tab:CSs}).

\subsubsection{Transformation matrices}
\label{appx:Ma-XYZ_Eq-E}

 Following the calculations
 done by A.~Bandyopadhyay and D.~Majumdar
 in Ref.~\cite{Bandyopadhyay10},
 the transformation matrices
 between the Equatorial
 and the Earth coordinate systems
 can be expressed by
 \cite{DMDDD-N}
\cheqnXa{A}
\beq
     \MaEqE(t)
  =  \left[\begin{array}{c c c}
             \gamma(t) \cos(\psiyear(t))                 ~&~
             \gamma(t) \sin(\psiyear(t)) \cos(\psiEarth) ~&~
              0                                           \\
            -\gamma(t) \sin(\psiyear(t)) \cos(\psiEarth) ~&~
             \gamma(t) \cos(\psiyear(t))                 ~&~
              0                                           \\
              0                                          ~&~
              0                                          ~&~
              1                                           \\
           \end{array}\right]
\~,
\label{eqn:Ma_Eq_E}
\eeq
 and,
 conversely,
\cheqnXb{A}
\beq
     \MaEEq(t)
  =  \left[\begin{array}{c c c}
             \gamma(t) \cos(\psiyear(t))                 ~&~
            -\gamma(t) \sin(\psiyear(t)) \cos(\psiEarth) ~&~
              0                                           \\
             \gamma(t) \sin(\psiyear(t)) \cos(\psiEarth) ~&~
             \gamma(t) \cos(\psiyear(t))                 ~&~
              0                                           \\
              0                                          ~&~
              0                                          ~&~
              1                                           \\
           \end{array}\right]
\~,
\label{eqn:Ma_E_Eq}
\eeq
\cheqnX{A}
 where we define
\beq
         \gamma(t)
 \equiv  \frac{1}
              {\sqrt{  \cos^2(\psiyear(t))
                     + \sin^2(\psiyear(t)) \cos^2(\psiEarth) } }
\~.
\label{eqn:psi_yr_Earth}
\eeq
\subsection{Laboratory--dependent coordinate systems}
\label{appx:XYZ_H-Lab}

 Now
 we come to
 the horizontal and the laboratory coordinate systems.

\subsubsection{Definitions}
\label{appx:def-XYZ_H-Lab}
\InsertSKPPlotD
 {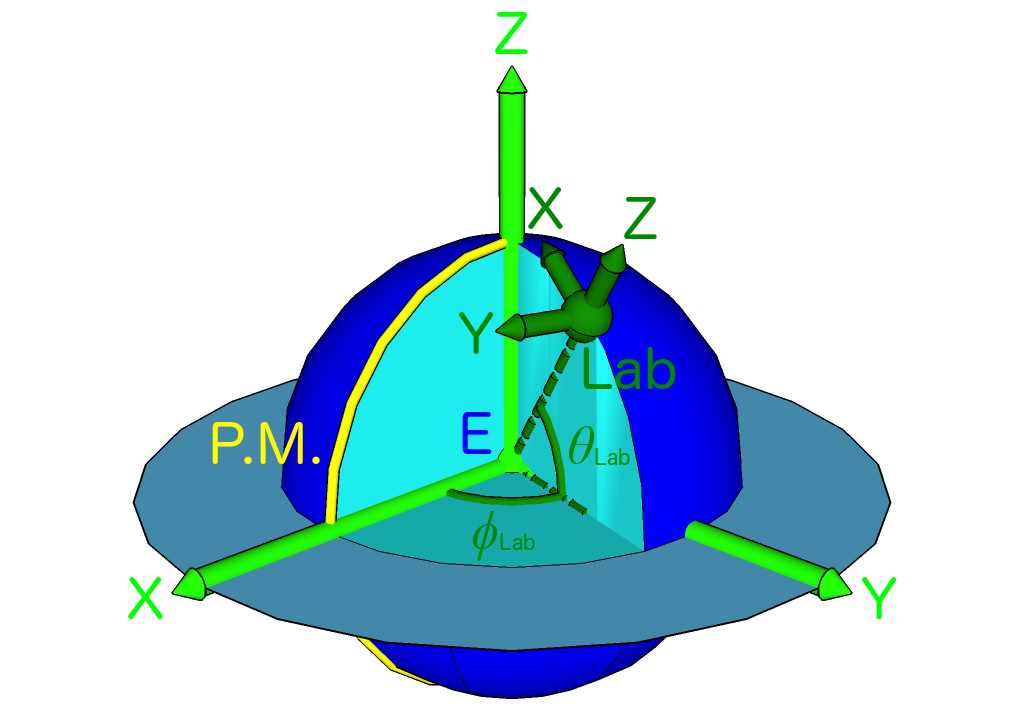}
 {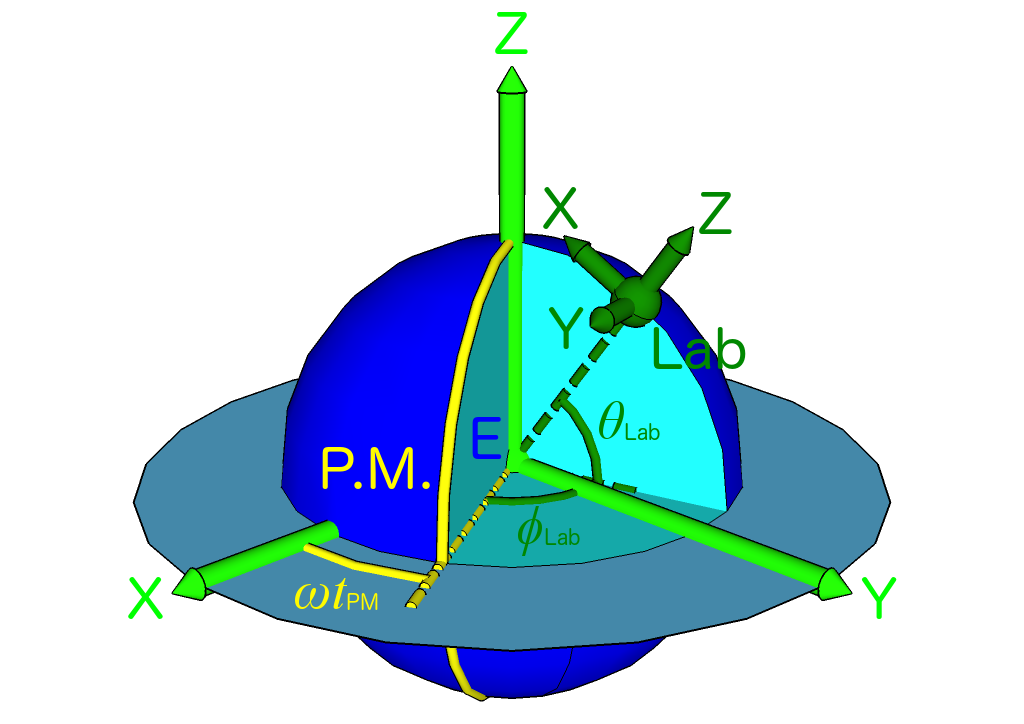}
 {fig:E-H-Lab}
 {The definitions of
  the (dark--green) horizontal (a)
  and laboratory (b) coordinate systems.
  $\phiLab$ and $\thetaLab$ indicate
  the longitude and the latitude of
  the location of the laboratory of interest,
  respectively.
  $\omega \tPM$ indicates
  the rotation angle of the (yellow) Prime Meridian
  from (UTC) 12 midnight (the beginning) of
  each single Solar day.
  As a reference,
  our (light--green)
  Earth coordinate system
  is also sketched here.
  See the text for the detailed descriptions.%
  }

 In Fig.~\ref{fig:E-H-Lab}
 we sketch
 the definitions of
 the (dark--green) horizontal (a)
 and laboratory (b) coordinate systems,
 respectively.
 Our (light--green)
 Earth coordinate system
 is also sketched here
 as a reference.

 At first,
 the origin of the horizontal coordinate system
 is chosen as the location
 of the laboratory of interest
 at (UTC) 12 midnight (the beginning) of
 each single Solar day
 with
 $\phiLab$ and $\thetaLab$ indicating
 the longitude and the latitude of
 the laboratory location,
 respectively.
 The primary direction (the $\xH$--axis)
 and the $\zH$--axis
 point towards north
 and the zenith,
 respectively.
 Then,
 as usual,
 the right--handed convention is used
 for defining the $\yH$--axis.
 Note that,
 as the Earth coordinate system,
 for each single (Solar) day,
 our horizontal coordinate system
 is fixed with the direction of the Prime Meridian
 at (UTC) 12 midnight
 and thus
 changes daily and discretely.

 Moreover,
 we consider also
 the instantaneous (UTC) incoming/scattering time of
 each recorded WIMP event
 and define
 our laboratory coordinate system.
 It is basically the same as the horizontal coordinate system,
 but rotates with the considered laboratory
 around the Earth's north polar ($\zEq/\zE$--axis)
 by an angle of $\omega \tPM$,
 where
\beq
         \omega
 \equiv  \frac{2 \pi}{1~{\rm day}}
\~,
\label{eqn:omega}
\eeq
 and
 $\tPM$ indicates
 the fractional part of
 the UTC incoming/scattering time $t$ of
 each recorded WIMP event
 in unit of day.
 Note that
 our laboratory coordinate system changes
 (rotates around the Earth's north polar axis)
 event by event
 (see Table \ref{tab:CSs}).

\subsubsection{Transformation matrices}
\label{appx:Ma-XYZ_H-Lab}

 At first,
 the transformation matrices
 between the Earth
 and the horizontal coordinate systems
 can be given by
 \cite{DMDDD-N}
\cheqnXa{A}
\beq
     \MaEH(\phiLab, \thetaLab)
  = {\small
     \left[\begin{array}{c c c}
            -\cos(\phiLab) \sin(\thetaLab) ~&~
            -\sin(\phiLab) \sin(\thetaLab) ~&~
                           \cos(\thetaLab)  \\
             \sin(\phiLab)                 ~&~
            -\cos(\phiLab)                 ~&~
              0                             \\
             \cos(\phiLab) \cos(\thetaLab) ~&~
             \sin(\phiLab) \cos(\thetaLab) ~&~
                           \sin(\thetaLab)  \\
           \end{array}\right]}
\~,
\label{eqn:Ma_E_H}
\eeq
 and then,
 conversely,
 we have
\cheqnXb{A}
\beq
     \MaHE(\phiLab, \thetaLab)
  = {\small
     \left[\begin{array}{c c c}
            -\cos(\phiLab) \sin(\thetaLab) ~&~
             \sin(\phiLab)                 ~&~
             \cos(\phiLab) \cos(\thetaLab)  \\
            -\sin(\phiLab) \sin(\thetaLab) ~&~
            -\cos(\phiLab)                 ~&~
             \sin(\phiLab) \cos(\thetaLab)  \\
                           \cos(\thetaLab) ~&~
              0                            ~&~
                           \sin(\thetaLab)  \\
           \end{array}\right]}
\~.
\label{eqn:Ma_H_E}
\eeq
\cheqnX{A}%
 Remind that,
 since,
 by our definitions,
 both of
 the Earth
 and the horizontal coordinate systems
 are fixed with the direction of the Prime Meridian
 at (UTC) 12 midnight of
 each single Solar day,
 the transformations
 between them
 depend only on the location
 (the longitude and the latitude)
 of the considered laboratory.

 Similarly
 the transformation matrices
 between the Earth
 and the laboratory coordinate systems
 can be obtained directly as
\cheqnXa{A}
\beqn
 \conti
     \MaELab(t, \phiLab, \thetaLab)
     \non\\
 \= {\footnotesize
     \left[\begin{array}{c c c}
            -\cos\abrac{\phiLab + \omega \tPM} \sin(\thetaLab) ~&~
            -\sin\abrac{\phiLab + \omega \tPM} \sin(\thetaLab) ~&~
                                               \cos(\thetaLab)  \\
             \sin\abrac{\phiLab + \omega \tPM}                 ~&~
            -\cos\abrac{\phiLab + \omega \tPM}                 ~&~
              0                                                 \\
             \cos\abrac{\phiLab + \omega \tPM} \cos(\thetaLab) ~&~
             \sin\abrac{\phiLab + \omega \tPM} \cos(\thetaLab) ~&~
                                               \sin(\thetaLab)  \\
           \end{array}\right]}
\~,
\label{eqn:Ma_E_Lab}
\eeqn
 and
\cheqnXb{A}
\beqn
 \conti
     \MaLabE(t, \phiLab, \thetaLab)
     \non\\
 \= {\footnotesize
     \left[\begin{array}{c c c}
            -\cos\abrac{\phiLab + \omega \tPM} \sin(\thetaLab) ~&~
             \sin\abrac{\phiLab + \omega \tPM}                 ~&~
             \cos\abrac{\phiLab + \omega \tPM} \cos(\thetaLab)  \\
            -\sin\abrac{\phiLab + \omega \tPM} \sin(\thetaLab) ~&~
            -\cos\abrac{\phiLab + \omega \tPM}                 ~&~
             \sin\abrac{\phiLab + \omega \tPM} \cos(\thetaLab)  \\
                                               \cos(\thetaLab) ~&~
              0                                                ~&~
                                               \sin(\thetaLab)  \\
           \end{array}\right]}
\~.
\label{eqn:Ma_Lab_E}
\eeqn
\cheqnX{A}%
 Then,
 by combining Eqs.~(\ref{eqn:Ma_H_E}) and (\ref{eqn:Ma_Lab_E})
 with Eqs.~(\ref{eqn:Ma_E_Lab}) and (\ref{eqn:Ma_E_H}),
 the transformation matrices
 between the horizontal
 and the laboratory coordinate systems
 can be expressed as
\cheqnXa{A}
\beqn
 \conti
     \MaHLab(t, \phiLab, \thetaLab)
     \non\\
 \=  \MaELab(t, \phiLab, \thetaLab) \~
     \MaHE(\phiLab, \thetaLab)
     \non\\
 \= {\scriptsize
     \left[\renewcommand{\arraystretch}{2}
           \begin{array}{c c c}
                      \cos(\omega \tPM)  \sin^2(\thetaLab) + \cos^2(\thetaLab) ~&~
                      \sin(\omega \tPM)  \sin  (\thetaLab)                     ~&~
            \bbig{1 - \cos(\omega \tPM)} \sin  (\thetaLab)   \cos  (\thetaLab)  \\
                    - \sin(\omega \tPM)  \sin  (\thetaLab)                     ~&~
                      \cos(\omega \tPM)                                        ~&~
                      \sin(\omega \tPM)                      \cos  (\thetaLab)  \\
            \bbig{1 - \cos(\omega \tPM)} \sin  (\thetaLab)   \cos  (\thetaLab) ~&~
                    - \sin(\omega \tPM)                      \cos  (\thetaLab) ~&~
                      \cos(\omega \tPM)  \cos^2(\thetaLab) + \sin^2(\thetaLab)  \\
           \end{array}\right]}
,
\label{eqn:Ma_H_Lab}
\eeqn
 and
\cheqnXb{A}
\beqn
 \conti
     \MaLabH(t, \phiLab, \thetaLab)
     \non\\
 \=  \MaEH(\phiLab, \thetaLab) \~
     \MaLabE(t, \phiLab, \thetaLab)
     \non\\
 \= {\scriptsize
     \left[\renewcommand{\arraystretch}{2}
           \begin{array}{c c c}
                      \cos(\omega \tPM)  \sin^2(\thetaLab) + \cos^2(\thetaLab) ~&~
                    - \sin(\omega \tPM)  \sin  (\thetaLab)                     ~&~
            \bbig{1 - \cos(\omega \tPM)} \sin  (\thetaLab)   \cos  (\thetaLab)  \\
                      \sin(\omega \tPM)  \sin  (\thetaLab)                     ~&~
                      \cos(\omega \tPM)                                        ~&~
                    - \sin(\omega \tPM)                      \cos  (\thetaLab)  \\
            \bbig{1 - \cos(\omega \tPM)} \sin  (\thetaLab)   \cos  (\thetaLab) ~&~
                      \sin(\omega \tPM)                      \cos  (\thetaLab) ~&~
                      \cos(\omega \tPM)  \cos^2(\thetaLab) + \sin^2(\thetaLab)  \\
           \end{array}\right]}
.
\label{eqn:Ma_Lab_H}
\eeqn
\cheqnX{A}%
 Note that,
 while
 the transformations
 between the Earth
 and the laboratory coordinate systems
 in Eqs.~(\ref{eqn:Ma_E_Lab}) to (\ref{eqn:Ma_Lab_E})
 depend on
 both of the laboratory location
 and
 the incoming/scattering time of
 each WIMP event
 (recorded in the considered laboratory),
 those
 between the horizontal
 and the laboratory coordinate systems
 are longitude ($\phiLab$) independent and
 depend only on
 the latitude of the laboratory location $\thetaLab$
 and the incoming/scattering time $t$.

\begin{table} [t!]
\small
\begin{center}
\renewcommand{\arraystretch}{1.5}
\begin{tabular}{|| c || c | c || c ||}
\hline
\hline
 \makebox[4   cm][c]{Coordinate system} &
 \makebox[2.5 cm][c]{Movement}          &
 \makebox[2.5 cm][c]{Rotation}          &
 \makebox[6   cm][c]{Style}             \\
\hline
\hline
 Galactic   & $\times$          & $\times^\dagger$ & Fixed \\
\hline
 Ecliptic   & $\surd$           & $\times$         & Orbital $\to$ approximately linear \\
 Equatorial & $\surd$           & $\times$         & Linear + orbital $\to$ spiral      \\
\hline
 Earth      & $\times^\ddagger$ & $\surd$          & Daily and discrete           \\
 Horizontal & $\times^\ddagger$ & $\surd$          & Daily and discrete           \\
 Laboratory & $\times^\ddagger$ & $\surd$          & Instantaneous and continuous \\
\hline
\hline
\end{tabular}
\end{center}
\caption{
 The summary of
 the styles of the movements and the rotations
 of all six celestial coordinate systems
 defined in our simulation package.
\\
 $^\dagger$:
 The tiny angle
 swept by the connection
 between the Solar and the Galactic centers
 during the orbital motion of the Solar system
 in the Galaxy
 is ignored in our package.
\\
 $^\ddagger$:
 Fixed on the Earth
 and combined additionally with
 the ``linear + orbital $\to$ spiral'' movement of
 the Equatorial coordinate system.
}
\label{tab:CSs}
\end{table}
\subsection{Incoming--WIMP coordinate system}
\label{appx:XYZ_chi}

 For our Monte Carlo simulation of
 3-D elastic WIMP--nucleus scattering events,
 we have introduced the ``incoming--WIMP'' coordinate system
 defined in Sec.~\ref{sec:XYZ_chi}.

\subsubsection{Transformation matrices}
\label{appx:Ma-XYZ_chi-Lab}

 Similar to the transformations
 between the horizontal
 and the Earth coordinate systems,
 the transformation
 from the incoming--WIMP coordinate system
 to the laboratory coordinate system
 can be done by rotating
 at first $\pi / 2 - \thetachiLab$ around the $\ychi$--axis
 and then $\pi     - \phichiLab$   around the $\zchi$--axis
 (see Fig.~\ref{fig:chi-Lab})
 and
 can thus be given by
\cheqnXa{A}
\beqn
 \conti
     \MachiLab(\phichiLab, \thetachiLab)
     \non\\
 \=  \left[\begin{array}{c c c}
            -\cos(\phichiLab) \sin(\thetachiLab) ~&~
             \sin(\phichiLab)                    ~&~
             \cos(\phichiLab) \cos(\thetachiLab)  \\
            -\sin(\phichiLab) \sin(\thetachiLab) ~&~
            -\cos(\phichiLab)                    ~&~
             \sin(\phichiLab) \cos(\thetachiLab)  \\
                              \cos(\thetachiLab) ~&~
              0                                  ~&~
                              \sin(\thetachiLab)  \\
           \end{array}\right]
\~.
\label{eqn:Ma_chi_Lab}
\eeqn
 Conversely,
 we also have
\cheqnXb{A}
\beqn
 \conti
     \MaLabchi(\phichiLab, \thetachiLab)
     \non\\
 \=  \left[\begin{array}{c c c}
            -\cos(\phichiLab) \sin(\thetachiLab) ~&~
            -\sin(\phichiLab) \sin(\thetachiLab) ~&~
                              \cos(\thetachiLab)  \\
             \sin(\phichiLab)                    ~&~
            -\cos(\phichiLab)                    ~&~
              0                                   \\
             \cos(\phichiLab) \cos(\thetachiLab) ~&~
             \sin(\phichiLab) \cos(\thetachiLab) ~&~
                              \sin(\thetachiLab)  \\
           \end{array}\right]
\~.
\label{eqn:Ma_Lab_chi}
\eeqn
\cheqnX{A}
%

%

%
%
%
 %
%

%
%

%

%
%
%
\end{document}